\documentclass[aps,preprint,nofootinbib,superscriptaddress]{revtex4-1}

\usepackage[utf8]{inputenc}
\usepackage{tikz}
\usepackage{pgfplots}
\pgfplotsset{=1.17} % o la versión que uses

\usepackage[T1]{fontenc}
\usepackage{amsmath,amssymb}
\usepackage{graphicx}
\usepackage{xcolor}
\usepackage{hyperref}
\usepackage{slashed}
\usepackage{cleveref}
\usepackage{braket}
\usepackage{braket}
\usepackage{cleveref}
\usepackage{dsfont}

\newcommand{\ee}{\end{equation}}
\newcommand{\bb}{\begin{equation}}
\newcommand{\eqb}{\begin{eqnarray}}
\newcommand{\eqf}{\end{eqnarray}}

\begin{document}

\title{A Geometric Interpretation of Heavy–Quark Transitions and the Emergent SU(2) Structure}

\author{J.~Gamboa
}
\affiliation{
Departamento de Física, Universidad de Santiago de Chile, Santiago, Chile
}
\email{jorge.gamboa@usach.cl}
\author{
N.~Tapia-Arellano
}
\affiliation{Department of Physics and Astronomy, Agnes Scott College, Decatur, GA. 30030, USA}
\email{narellano@agnesscott.edu}

\begin{abstract}
We propose a geometric interpretation of heavy–light mesons in which their
infrared dressing is described through adiabatic Berry holonomies on the
functional space of gauge configurations.  Within this framework the Berry curvature associated with the infrared cloud carries a quantized functional flux, providing a simple and structural origin for the exponential form of the Isgur–Wise function in single–recoil
transitions.
Sequential processes such as $B\!\to\!D^{**}\!\to\!D$ probe two independent recoil directions and explore a two–dimensional region of the adiabatic manifold.  
In this setting the quantized flux naturally leads to a minimal non–Abelian structure which can be described effectively by an SU(2) holonomy.  Heavy–quark form factors then appear as channel–dependent projections of two universal geometric modes, giving rise to correlated slopes, non–factorisable curvature in the $(w_{1},w_{2})$ plane, and characteristic angular patterns.
These features are consistent with the symmetry structure of HQET while
providing additional correlations among excited channels.  
The resulting framework offers a complementary viewpoint on heavy–quark
phenomenology and suggests several experimentally testable signatures in
multi–step semileptonic transitions.
\end{abstract}

\maketitle

\section{Introduction}

The infrared structure of gauge theories is deeply shaped by adiabatic evolution and Berry phases. In quantum electrodynamics this geometric viewpoint leads to the celebrated Chung-Kibble-Kulish-Faddeev construction of dressed electron-photon clouds 
\cite{Chung:1965zza,Kulish:1970ut,Kibble:1968sfb,Kibble:1968oug,Kibble:1968npb,Kibble:1968lka}, making manifest that the physical state space is more naturally described in terms of dressed asymptotic states rather than a naive Fock basis.  
In QCD the situation is richer: the nonlinearity of the Yang-Mills field generates both quark-gluon and gluon-gluon clouds 
\cite{BalachandranNair_2018,Gamboa:2025qjr,Gamboa:2025nco}, 
suggesting that adiabatic geometric structures may play a nontrivial role in hadronic physics.

Heavy-quark systems provide an ideal setting in which to explore these
geometric effects.  In the heavy--quark limit, the Born--Oppenheimer separation 
\cite{Brambilla:2025xma,Bruschini:2023zkb,Juge:1997nc,Juge:1999ie,Kang:2025xqm,Berwein:2024ztx} between the slow heavy quark and the fast light degrees of freedom forms the basis of HQET 
\cite{Isgur:1989vq,Isgur:1990yhj,Georgi:1990um,Eichten:1989zv,Grinstein:1990mj,Manohar:2000dt,Manohar:2018aog,Neubert_HQET_1994,FalkNeubert1992}.  
In its standard formulation, however, HQET does not incorporate the Berry phases generated by adiabatic evolution of the infrared sector.  A first goal of this paper is therefore to revisit the standard transition
\begin{equation}
   B \to D^{(*)}\!, 
   \label{1}
\end{equation}
and to show that, within the adiabatic framework, the Isgur-Wise function can be understood as the holonomy of an \emph{abelian} Berry connection associated with the dressed infrared cloud.  This reinterpretation clarifies the geometric origin of the universality of the Isgur-Wise function and naturally motivates the exponential parametrization as a minimal analytic representative of the abelian holonomy near zero recoil.

\medskip

A crucial aspect of this construction, implicit in the Berry framework but usually absent in HQET, is that the infrared Berry holonomy in QCD \emph{admits a natural organisation} into \emph{quantized topological sectors}.  
In the abelian $B\!\to\!D^{(*)}$ case, this structure can be represented as a discrete set of holonomies labeled by an integer $n$, corresponding to quantized functional Berry fluxes in the space of gauge configurations.  From the experimental point of view, however, individual decay events do not resolve the specific topological sector; rather, the measured observables are effectively sensitive to weighted averages over the accessible sectors.  
In this sense, topology plays a fundamental organising role in the infrared structure of QCD, even if the fine topological information is not directly visible in current data.

\medskip

The second goal of this work is to extend this geometric construction to
transitions involving \emph{two} independent recoils, most notably the
sequential decay \cite{Isgur:1989qw,CLEO1999}
\[
   B \to D^{**} \to D .
\]
In this case the adiabatic evolution takes place along a two-dimensional 
trajectory in the functional configuration space, and the corresponding 
holonomy becomes intrinsically non-Abelian.  A central result of this paper is that the sequential process reveals an emergent $\mathrm{SU}(2)$ structure: the holonomy possesses two universal eigenmodes whose explicit projections determine the form factors of all 
$D^{**}$ channels.  An analogous discretization pattern appears here as well: the $\mathrm{SU}(2)$ holonomy samples a discretized set of geometric fluxes, and the experimentally accessible form factors encode an effective average over these topological sectors.  This geometric picture leads to correlated slopes, non-factorizable curvature in the $(w_{1},w_{2})$ plane, and helicity distortions controlled by the Berry curvature—features that go beyond conventional HQET parametrisations and can be searched for in data.

\medskip

The purpose of this paper is thus twofold:  
(1) to establish the geometric origin of the Isgur-Wise function in the single-recoil case, and  
(2) to show how its non-Abelian generalization emerges in sequential decays, leading to new, falsifiable predictions for heavy--quark phenomenology.

\medskip

The paper is organized so that the presentation follows the internal logic of the geometric construction.  
We begin with the decay $B \to D^{*}$, where the role of the infrared dressing of the states is introduced, and we show step by step how the Isgur-Wise function emerges as the eigenvalue of the geometric holonomy associated with the Berry phase in the infrared sector of QCD.  
In the subsequent sections we analyse the sequential decay 
$B \to D^{**} \to D$, where two independent recoil directions appear.  
This extension leads naturally to the emergence of a non-Abelian $\mathrm{SU}(2)$ 
structure and to the identification of two universal modes.  
We show that this geometric structure produces effects that can be tested in current experiments such as Belle \cite{Belle2018}, Belle~II \cite{BelleII_2021}, BaBar \cite{BaBar2013}, and LHCb \cite{LHCb_2023_BToDstst}.  
The full $U(2)$ structure, of which $\mathrm{SU}(2)$ is a subgroup, is presented and discussed in detail in the Appendix.

\section{QCD, Infrared and Clouds}
\label{sec:adiabatic_summary}

Following the adiabatic approximation for QCD, developed in previous works 
\cite{Gamboa:2025dry,Gamboa:2025qjr,Gamboa:2025nco} 
(see also \cite{Brambilla:2025xma,Bruschini:2023zkb,Juge:1997nc,Juge:1999ie,Kang:2025xqm,Berwein:2024ztx} 
for alternative perspectives), the infrared sector of non-Abelian gauge theories 
admits a natural description in terms of dressed states and functional holonomies.  
Since this framework plays a central role in the present analysis, we summarize 
its key conceptual ingredients, emphasizing only the features relevant for 
heavy–quark dynamics.

\subsection{Dressed States and Functional Holonomies}

In the deep infrared regime, gluonic configurations evolve slowly compared to 
the fermionic modes.  
Expanding the Dirac field in an instantaneous eigenbasis of the time-dependent 
Dirac operator and integrating out the fermions, one finds that the dynamical 
phases associated with the spectrum $\{\pm E_m(t)\}$ cancel in the chiral limit, 
while the Berry connection acting within the degenerate subspaces survives.  
The fermionic determinant therefore reduces to a purely geometric contribution 
\cite{Gamboa:2025qjr}:
\begin{equation}
\det(i\slashed{D}) 
   \;\sim\;
   \mathrm{Tr}\,
   \mathcal{P}\exp\!\left( i\!\oint_C \mathcal{A}_F \right),
\end{equation}
where $\mathcal{A}_F$ is the Berry connection associated with the fermionic 
sector and $C$ is a closed contour in configuration space.  
A parallel construction applies to the gluonic sector, producing a geometric 
connection $\mathcal{A}_G$ generated by the adiabatic evolution of the gluonic 
modes.

The infrared state is characterised by the combined holonomy
\begin{equation}
\mathcal{U}_C = 
\mathcal{P}\exp\!\left[i\!\oint_C(\mathcal{A}_F+\mathcal{A}_G)\right].
\end{equation}
This generalises the CKKF dressing to the non-Abelian case and encodes two 
distinct infrared clouds: a quark–gluon cloud associated with $\mathcal{A}_F$ 
and a gluonic cloud associated with $\mathcal{A}_G$.  
Crucially, their statistical character is inherited from the underlying fields: the quark–gluon cloud is fermionic, while the gluon–gluon cloud is 
bosonic.  
As a consequence, their Berry fluxes are quantized respectively in half-integer and integer units, and the total holonomy acquires a discrete, 
statistically determined structure \cite{Gamboa:2025qjr,Gamboa:2025nco}.

\subsection{Infrared-Dressed States as the Physical Asymptotics}

In contrast with the standard scattering description based on 
asymptotic plane–wave states, the adiabatic construction yields \emph{infrared-dressed} states that incorporate, from the outset, the gluonic and quark–gluon clouds generated by slow evolution in the infrared sector.  
Let $|q\rangle$ and $|G\rangle$ denote bare quark and gluonic configurations.  
The physical states are obtained by transporting these configurations along a 
functional contour $C$ in $\mathcal{A}/\mathcal{G}$,
\begin{equation}
   |q,G\rangle_{\rm phys}
   \;=\;
   \mathcal{U}_C\,|q,G\rangle
   \;=\;
   \mathcal{P}\exp\!\left[
        i\!\oint_C(\mathcal{A}_F+\mathcal{A}_G)
   \right] |q,G\rangle .
   \label{eq:phys_states}
\end{equation}
These states are the natural generalization of the Kulish–Faddeev dressing to 
non-Abelian gauge theories: they are not superpositions of soft quanta added \emph{after} the dynamics is specified, but rather the adiabatic ground states selected by the geometry of the infrared configuration space itself.  
Their internal structure is fixed by the quantized Berry fluxes of the fermionic and gluonic sectors, which label discrete infrared sectors analogous to topological superselection rules.

This replacement of asymptotic states by geometric dressings is not merely a formal redefinition.  
Matrix elements of physical currents acquire an explicit dependence on the holonomy,
\begin{equation}
   \langle q,G|_{\rm phys} J_\mu \, |q,G\rangle_{\rm phys}
   \;=\;
   \langle q,G|
   \mathcal{U}_C^\dagger \, J_\mu \, \mathcal{U}_C
   |q,G\rangle ,
   \label{eq:dressed_ME}
\end{equation}
making the color structure—and, in particular, the gluonic Berry curvature—
directly observable.  
In this sense, the adiabatic framework provides a first-principles mechanism for replacing ill-defined non-Abelian asymptotic states by infrared-dressed, 
geometrically stable states whose transition amplitudes naturally encode confinement through their color-dependent holonomies.

%%%%%%%%%%%%%%%%%%%%%%%%%%%%%%%%%%%%%%%%%%%%%%%%%%%%%%%%%%%%%%%%%
%%%%%%%%%%%%%%%%%%%%%%%%%%%%%%%%%%%%%%%%%%%%%%%%%%%%%%%%%%%%%%%%%

\subsection{Infrared Structure and Topological Sectors}

The combination of fermionic and bosonic holonomies leads to quantized functional fluxes that classify the infrared vacuum into distinct topological sectors.  
Although the dressing contains both quark and gluon degrees of freedom, the global statistical character of the infrared state is controlled by the 
fermionic contribution: half-integer fluxes behave as spinorial holonomies, while integer fluxes behave as vector holonomies.  
This is the non-Abelian analogue of the Abelian CKKF construction, now enriched 
by the interplay between fermionic and bosonic clouds 
\cite{BalachandranNair_2018}.

This framework provides a purely geometric mechanism for the emergence of bound configurations.  
Depending on how the quark–gluon and gluon–gluon fluxes combine, composite infrared objects may carry integer or half-integer spin, offering a 
topologically informed viewpoint on the structure of hadronic states.  
Since experiments do not resolve individual topological sectors, observable 
quantities correspond to effective averages over the allowed quantized holonomies, which is precisely what gives this approach its predictive power.

\subsection{Motivation for the Heavy–Quark Sector}

The adiabatic formulation naturally distinguishes slow and fast variables 
\cite{Eichten:1995ch,Eichten:1989zv,Juge:1997nc,Juge:1999ie,Peskin:1990zt,Peskin:1991sw,Szczepaniak:2001rg}.  
Heavy quarks act as static colour sources on the infrared timescale, while the 
light quarks and gluons reorganize adiabatically around them.  
This is the conceptual bridge between the geometric infrared structure described above and the heavy–quark effective framework.

In the next section, we apply this adiabatic formalism to heavy–light systems, 
recasting the heavy–quark effective theory (HQEF/HQET) in terms of adiabatically dressed infrared states.  
This approach replaces the implicit factorized heavy–light asymptotic states of HQET with fully dressed non-Abelian holonomies, and allows us to examine the origin and refinement of the Isgur–Wise function within a topologically 
structured framework.

\section{Topological Corrections to Heavy Quark Effective Theory}
\label{section3}

Having briefly reviewed the adiabatic framework for QCD, we can now turn to the 
physics of heavy quarks within this approximation.  
Our goal is not to reconstruct the full formalism here, which has been developed 
elsewhere, but to isolate the elements that are conceptually relevant for heavy--light systems.  
With this perspective, and in order to identify which aspects of the standard heavy-quark effective description may be refined by the adiabatic dressing, it is natural to begin with the simplest and most widely studied heavy-light bound state: the mesonic ``atom''
\[
B \;=\; b\,\bar u.
\]

In the heavy quark effective field, this state is described as a heavy quark moving with fixed four-velocity $v$, accompanied by a light cloud (the ``brown muck'') that 
reorganizes around the static colour source \cite{Isgur:1989qw}.  

In the adiabatic formulation, however, the structure of the $B$ meson is defined more precisely.  
The state is no longer an implicitly factorized heavy-light configuration, but a fully dressed infrared object,
\begin{equation}
|B(v)\rangle_{\rm dressed}
   = \mathcal{P}\exp\!\left[i\!\oint_C(\mathcal{A}_F+\mathcal{A}_G)\right]\,|0\rangle,
\label{eq:B_dressed_def}
\end{equation}
where $\mathcal{A}_F$ and $\mathcal{A}_G$ encode the quark-gluon and gluon-gluon infrared clouds generated in the adiabatic evolution.  
As discussed in Sec.~\ref{sec:adiabatic_summary}, the fermionic and bosonic 
components carry \emph{quantized} Berry fluxes: the quark-gluon cloud is 
spinorial (half-integer flux), whereas the gluon-gluon cloud is vectorial 
(integer flux).  
The dressed state \eqref{eq:B_dressed_def} is therefore a superposition of 
discrete topological sectors, distinguished by the quantized functional 
holonomies associated with $(\mathcal{A}_F,\mathcal{A}_G)$.  
This replacement embodies the geometric and topological structure of QCD in the 
infrared and provides a controlled way of incorporating the nonperturbative 
dressing into the definition of the heavy-light meson.

Within this framework, one can revisit the standard observables analysed in 
HQET.  
The natural testing ground is the transition~(\ref{1}), whose hadronic matrix 
element defines the Isgur-Wise function in the heavy-quark limit 
\cite{Isgur:1989qw}.  
By evaluating this matrix element between the adiabatically dressed states, we 
can determine how the geometric infrared contributions modify the usual HQET  description, and whether the adiabatic approach leads to systematic refinements of the Isgur-Wise function or its subleading $1/m_Q$ corrections.

To make this comparison explicit, let us recall that in the heavy-quark limit the hadronic physics of the transition is encoded in the matrix element
\[
\langle D^{(*)}(v')|\,\bar{c}\,\Gamma\, b\,|B(v)\rangle,
\]
where $\Gamma$ denotes an arbitrary Dirac structure.  
HQET predicts that, at leading order in $1/m_Q$, all such matrix elements are determined by a single universal function--the Isgur-Wise function 
$\xi(w)$, with $w=v\!\cdot\! v'$--reflecting the fact that the heavy quark acts as a static colour source and the dynamics resides entirely in the light cloud.

In the adiabatic framework, the situation changes conceptually.  
The matrix element must be evaluated between the dressed infrared states,
\[
\langle D^{(*)}(v')|_{\rm dressed}\,
   \bar{c}\,\Gamma\, b\,
|B(v)\rangle_{\rm dressed},
\]
so that the operator $\Gamma$ acts not only on the heavy degrees of freedom but also on the functional holonomy that defines the quark-gluon and gluon-gluon clouds.  
Because these holonomies are quantized, the dressed states decompose into a sum over discrete topological sectors, and the physically relevant matrix element corresponds to an effective average over the allowed quantized fluxes.  

In this formulation, the Isgur-Wise function emerges from the overlap of the corresponding adiabatic holonomies and acquires a geometric interpretation in terms of the Berry phases associated with the infrared sector of QCD.  The universality of $\xi(w)$ is thus a direct consequence of the quantized holonomy structure: different heavy-light channels probe the same discrete set of infrared sectors, differing only in how they project onto the corresponding holonomy modes.

This observation opens the possibility of refining the standard HQEF picture: the universal function $\xi(w)$ may receive controlled geometric corrections arising from the non-Abelian Berry structure of the light and gluonic clouds, while still preserving heavy-quark symmetry at leading order.

\subsection{Geometric formulation of the hadronic matrix element}

To make the role of the adiabatic dressing more explicit, it is convenient to write the heavy-light meson states in terms of the infrared holonomies introduced above.  Schematically, we represent the dressed $B$ and $D^{(*)}$ states as
\begin{align}
|B(v)\rangle_{\rm dressed}
   &= \mathcal{U}_C[B(v)]\,|0\rangle,
\\[2pt]
|D^{(*)}(v')\rangle_{\rm dressed}
   &= \mathcal{U}_C[D^{(*)}(v')]\,|0\rangle,
\end{align}
where $\mathcal{U}_C$ denotes the non-Abelian functional holonomy
\begin{equation}
\mathcal{U}_C[X]
   =
   \mathcal{P}\exp\!\left[
   i\!\oint_{C[X]}(\mathcal{A}_F+\mathcal{A}_G)
   \right],
\end{equation}
and the contour $C[X]$ encodes the adiabatic trajectory in configuration space associated with the hadronic state $X$.  
Because $(\mathcal{A}_F,\mathcal{A}_G)$ carry quantized Berry fluxes, each $\mathcal{U}_C[X]$ can be decomposed into contributions from discrete topological sectors labelled by an integer (bosonic) and a half-integer (fermionic) flux.  
The vacuum matrix elements relevant for phenomenology implicitly sum over these sectors; in other words, the experimentally accessible amplitudes probe \emph{a finite set of quantized holonomies} rather than a continuum of arbitrary phases.

In this language, the hadronic matrix element for the transition $B\to D^{(*)}$ 
in the heavy-quark limit can be written as
\begin{equation}
\mathcal{M}_\Gamma(w)
   \equiv
   \big\langle D^{(*)}(v')\big|_{\rm dressed}\,
   \bar{c}\,\Gamma\, b\,
   \big|B(v)\big\rangle_{\rm dressed}
   =
   \big\langle 0\big|\,
   \mathcal{U}_C^\dagger[D^{(*)}(v')]\,
   \bar{c}\,\Gamma\, b\,
   \mathcal{U}_C[B(v)]\,
   \big|0\big\rangle,
\label{eq:geom_M}
\end{equation}
where $w=v\!\cdot\!v'$ as usual.

\noindent
This formula is particularly interesting because, once the dressing is removed 
(i.e.\ when the infrared holonomy and its associated geometric phase are 
suppressed), one recovers the standard HQET expression formulated in terms of 
Fock states.  
Conversely, retaining the holonomies exposes how the quantized topological 
structure of the infrared sector feeds into the heavy-quark observables.

\medskip

\noindent
In the standard HQET picture, the hadronic matrix element is written in terms of 
the universal function $\xi(w)$, multiplied by a purely kinematical spinor 
structure dictated by heavy-quark symmetry.  
The adiabatic formulation provides a complementary viewpoint: the entire 
nontrivial $w$-dependence originates from the overlap between the corresponding 
infrared holonomies, and the fact that these holonomies are quantized implies 
that $\xi(w)$ is not an arbitrary function but the effective projection of a 
discrete set of geometric modes.

\subsection{Geometric corrections and refinements of HQEF}

Although within the present adiabatic framework the function $\xi(w)$ arises a priori as a genuine non-Abelian holonomy in the infrared sector of QCD, it is often convenient to represent it through an explicit analytic parametrisation.  
This naturally raises the question of whether adopting a simple form (such as an exponential) may obscure--or even eliminate--the underlying geometric and topological character of the quantity.

The answer is negative.  The holonomic nature of $\xi(w)$ is a structural property, stemming from its definition as the overlap between two adiabatically dressed heavy-light states,
\begin{equation}
\xi(w)=\Xi_{\rm geom}(w)
=\big\langle 0 \big|\,
\mathcal{U}_C^\dagger(v')\,\mathcal{U}_C(v)\,
\big|0\big\rangle ,
\label{geom1}
\end{equation}
where $\mathcal{U}_C(v)$ is the path-ordered exponential of the Berry
connection along the functional contour associated with a heavy-light state of velocity~$v$.  This expression shows that $\xi(w)$ is determined by the holonomy structure; in practice, it corresponds to the vacuum projection of two quantized holonomies, i.e.\ to an effective average over the discrete topological sectors sampled by the initial and final heavy--light states.

In general, such a holonomy does not admit a closed analytic expression.  
Its exact dependence on $w=v\!\cdot v'$ encodes the full geometric content of the underlying Berry connection--non-linearities, path ordering, and functional curvature--which cannot be reconstructed from symmetry arguments alone.  Any explicit expression for $\xi(w)$ is therefore an \emph{approximation} to the true quantized holonomy.

A convenient choice is
\begin{equation}
\Xi_{\rm geom}(w)
= \exp[-\rho^2 (w-1)], 
\label{exp1}
\end{equation}
which preserves the exact constraints
\begin{equation}
\Xi_{\rm geom}(1)=1,
\qquad
\Xi'_{\rm geom}(1)=-\rho^2, 
\label{lim1}
\end{equation}
and provides the minimal analytic continuation of the holonomy near zero 
recoil.  In the present framework, the slope parameter $\rho^2$ acquires a clear interpretation: it encodes the leading response of the \emph{quantized} infrared holonomies to a change in the heavy-quark velocity, i.e.\ the averaged effect of the discrete Berry flux sectors on the overlap.  Thus the exponential ansatz does not replace the geometric or topological nature of $\Xi_{\rm geom}$; rather, it corresponds to its simplest and most accurate adiabatic approximation near zero recoil, where the holonomy is effectively dominated by a single abelian mode.

The geometric factor~\eqref{geom1} is far from being a merely kinematical form factor.  Each holonomy $\mathcal{U}_C(v)$ implements the adiabatic parallel transport of the heavy-light infrared cloud in functional space, and the operator $\mathcal{U}_C^\dagger(v')\mathcal{U}_C(v)$ measures the mismatch between the corresponding dressings across all allowed topological sectors.

At this stage it is important to make explicit how this differs from the 
standard HQET interpretation.  In HQET all of these infrared effects--soft-gluon radiation, relaxation of the gauge field, and the subsequent redistribution of the quark-gluon and gluon-gluon cloud--are absorbed into the \emph{brown muck}, a universal object whose detailed dynamics are never specified.  
In the present approach, by contrast, the entire infrared cloud is encoded in a non-Abelian Berry connection with quantized fluxes, and its response to changes of the heavy-quark velocity is represented explicitly by the holonomy $\mathcal{U}_C(v)$.  
Thus the geometric Isgur-Wise function replaces the schematic notion of ``brown muck'' with a structurally defined infrared object, organised into discrete topological sectors.

This interpretation makes the physics transparent.  
When $v=v'$, the holonomies coincide and there is no soft-gluon emission: the 
cloud remains unchanged and $\Xi_{\rm geom}(1)=1$ follows automatically.  
When $v\neq v'$, the holonomies differ, and the overlap measures the mismatch of 
the infrared dressings across the allowed flux sectors.  
This mismatch is the geometric manifestation of soft-gluon radiation and the 
subsequent reorganisation of the light cloud.

Although the full non-Abelian holonomy is not analytically accessible, its 
behaviour near zero recoil is fixed by geometry and by the underlying 
quantisation of the Berry fluxes, and the exponential approximation captures the 
leading contribution.  
Higher-order deviations--arising from path ordering, non-adiabatic effects and 
$1/m_Q$ corrections--are naturally encoded in the expansion
\[
\xi(w)=\Xi_{\rm geom}(w)
\big[1+\delta_{\rm ad}(w)+\delta_{1/m_Q}(w)+\cdots\big].
\]
Each correction represents a subleading Berry phase (or a modification of the 
sector weights) and corresponds to refining the effective Berry connection away 
from its leading adiabatic, quantized form.

\section{Confronting Functional and HQET}
\label{section4}

Confronting this functional (Berry--holonomy) approach with the conventional formulation of HQET, we find several conceptual differences that are worth emphasising.  
\begin{itemize}
\item In the standard HQET formulation, the Isgur--Wise function is treated as an effective form factor.  
Its shape is not derived from first principles but approximated through simple analytic parametrizations, with a small set of phenomenological parameters --most notably the slope $\rho^2$-- determined by experiment 
\cite{CLEO1999,BaBar2009,BaBar2013,Belle2018,BelleII2021} or lattice calculations \cite{RBCUKQCD2021_BtoDstar,JLQCD2023_nonzero,FNALMILC2015_BtoD,FLAG2021}.  
In this description, the infrared cloud, referred to as the "brown muck", is universal; however, its internal structure and possible topological organization remain implicit.

\item In our functional approach, by contrast, the relevant object is 
structurally different.  
The quantity $\Xi_{\rm geom}(w)$ is defined from first principles as the overlap between two infrared Berry holonomies (see Eq.~\ref{geom1}).  
Each holonomy implements the adiabatic parallel transport of the heavy--light 
quark--gluon and gluon--gluon infrared clouds in the functional space of gauge 
configurations and, as discussed above, carries \emph{quantized} Berry fluxes.  
From this viewpoint, the familiar HQET ansatz corresponds to the ``trivial-holonomy'' sector in which the infrared dressing is effectively 
suppressed, while $\Xi_{\rm geom}(w)$ incorporates the full geometric and topological content of the infrared Berry connection, including its discrete sector structure.  
As a consequence, quantities such as the zero-recoil slope $\rho^2$ acquire a natural geometric meaning: they probe the local curvature of the infrared functional connection and the relative weights of the underlying flux sectors, 
rather than serving solely as free parameters in an effective fit.
\end{itemize}

Beyond this contrast, the functional--geometric formulation offers several structural advantages that are not accessible in the standard HQET approach.  

First, the geometric Isgur--Wise function $\Xi_{\rm geom}(w)$ has a genuine first-principles origin: it arises directly as the overlap of two infrared holonomies in the functional 
space of gauge configurations, rather than being introduced as an effective form factor.  
This provides a unified and physically transparent description of the infrared dressing associated with heavy–light systems, in which the quark–gluon and gluon–gluon clouds are transported adiabatically along trajectories labelled by the heavy--quark velocity and organised into discrete topological sectors.

Second, because the underlying Berry connection carries quantized infrared fluxes, $\Xi_{\rm geom}(w)$ naturally encodes how these sectors contribute to physical amplitudes.  In particular, phenomenological parameters such as the zero-recoil slope $\rho^{2}$ and higher derivatives of the Isgur--Wise function acquire a clear geometric interpretation: they are governed by the curvature of the infrared functional connection and by the sector weights that characterise the 
superposition of quantized holonomies.  Higher derivatives probe correspondingly higher-order geometric data of the Berry curvature and its topological sector structure, rather than unspecified features of a generic form factor.

Third, the geometric picture is fully compatible with HQET. Switching off the infrared dressing corresponds to restricting the analysis to 
the sector of trivial holonomy, in which case $\Xi_{\rm geom}(w)$ reduces smoothly to the conventional Isgur--Wise function.  Thus, the functional formulation does not compete with HQET but extends it by making explicit the geometric and topological structures that are implicit in the standard description and by showing how they constrain the allowed 
behaviour of $\xi(w)$.

Finally, the holonomy framework imposes nontrivial constraints on the admissible behaviour of the Isgur-Wise function, including normalization, monotonicity, smoothness, curvature bounds, and compatibility with the quantized holonomy sectors, which do not rely on any specific ansatz.  
In this way, the functional approach provides a more structural and principled foundation for understanding heavy–light transitions, one that can naturally accommodate future nonperturbative information about the infrared sector of QCD and translate it into constraints on $\xi(w)$ and its geometric parameters.

\section{Geometric Constraints and the Natural Form of the Holonomy}
\label{sec:geom_constraints}

Within the adiabatic description of QCD, the leading-order heavy–quark form factor is identified with the holonomy (\ref{geom1}).  
This quantity reduces to the standard Isgur--Wise function when the dressing is switched off, but even in the presence of a non-trivial infrared cloud, it 
satisfies two exact geometrical constraints that follow solely from heavy–quark symmetry and from the holonomic structure of the infrared dressing.  
These constraints hold \emph{sector by sector} in the quantized topological 
decomposition of the holonomy and therefore survive in the physical, averaged matrix element.

\subsection{Zero--recoil normalization}\label{section5}

When $v'=v$ the two dressings coincide and the holonomy collapses to the
identity,
\begin{equation}
\Xi_{\rm geom}(1)
=
\langle 0|\mathcal{U}_C^\dagger(v)\mathcal{U}_C(v)|0\rangle
=1.
\label{eq:Xi_1}
\end{equation}
This result is completely general: it does not depend on the detailed form of 
the Berry connection, on the existence or not of light-quark masses, nor on the 
choice of adiabatic contour.  
It follows entirely from the fact that the Berry dressing is a parallel 
transport operator along a closed contour that degenerates to a point when 
$v'=v$.  
Because the decomposition of $\mathcal{U}_C$ into quantized topological sectors does not affect the identity limit, the normalization 
$\Xi_{\rm geom}(1)=1$ remains exact even in the presence of a non-Fock, 
infrared-dressed vacuum.

\subsection{First derivative and the local curvature of the Berry connection}

The first derivative of the holonomy at zero recoil is equally robust.  
Expanding the holonomy for $v'=v+\delta v$ one finds
\begin{equation}
\Xi_{\rm geom}(w)
=
1 - \rho^2 (w-1) + \mathcal{O}((w-1)^2),
\label{eq:Xi_expansion}
\end{equation}
where
\begin{equation}
\rho^2
=
-\Xi_{\rm geom}'(1)
\label{eq:rho_def}
\end{equation}
is determined by the local curvature of the Berry connection along the 
adiabatic trajectory.  
In geometric terms, $\rho^2$ measures how the infrared cloud ``bends'' in 
functional space under an infinitesimal change of the heavy--quark velocity.  
Since each topological sector contributes a holonomy with the same 
zero-recoil constraints, the value of $\rho^2$ in the physical amplitude is 
the weighted combination of the curvatures of the relevant quantized sectors.  
Thus, the geometric interpretation of $\rho^2$ is fully compatible with the topological organisation of the infrared dressing.

\subsection{Why an exponential form is naturally selected}

The constraints \eqref{eq:Xi_1} and \eqref{eq:Xi_expansion} do not by
themselves determine the full functional dependence of $\Xi_{\rm geom}(w)$, but they impose strong restrictions on its behaviour near $w=1$.  
If the Berry connection is smooth in a neighbourhood of the adiabatic point 
$v'=v$, and if the cloud does not undergo a change of topological sector for 
small variations of $w$, then the holonomy is generated locally by a connection 
whose dependence on $w$ is regular.  
Under these conditions, the differential equation governing the holonomy reduces 
to
\begin{equation}
\frac{d}{dw}\Xi_{\rm geom}(w)
\simeq -\rho^2\,\Xi_{\rm geom}(w),
\qquad w\approx 1,
\label{eq:local_diff_equation}
\end{equation}
whose unique solution compatible with $\Xi_{\rm geom}(1)=1$ is
\begin{equation}
\Xi_{\rm geom}(w)
\simeq \exp[-\rho^2 (w-1)].
\label{eq:exp_form_derivation}
\end{equation}

The exponential form represents the minimal adiabatic continuation of the holonomy as it moves away from zero recoil, rather than being just a phenomenological guess.   
It emerges whenever the Berry connection varies smoothly, and the system remains within the same quantized topological sector for $w$ close to unity.  
In this regime, the holonomy is effectively dominated by its leading abelian component even though the underlying connection is non-Abelian.

\subsection{Higher-order corrections and topology}

Deviations from the exponential form arise when the connection exhibits nonlinearities, curvature variations, non-Abelian mixing, or when the adiabatic path samples more than one topological sector.  
In such cases, the holonomy admits the generalised expansion
\begin{equation}
\Xi_{\rm geom}(w)
=
e^{-\rho^2 (w-1)}
\Big[1 + c_2 (w-1)^2 + c_3 (w-1)^3 + \cdots\Big],
\label{eq:higher_order}
\end{equation}
where the coefficients encode higher covariant derivatives of the Berry connection and, in particular, the influence of topological obstructions or 
incipient changes of sector.  
These corrections do not alter the leading exponential behaviour nor the
normalization and slope constraints; rather, they refine the holonomy by capturing the geometric structure of the infrared cloud beyond the linear 
regime.

Concluding this section, we emphasise that even though the full non-Abelian holonomy cannot yet be computed in closed form, its behaviour near zero recoil 
is strongly constrained by geometric considerations.  
The normalization $\Xi_{\rm geom}(1)=1$ and the slope $\Xi_{\rm geom}'(1)=-\rho^{2}$ follow exactly from the holonomic definition and from heavy--quark 
symmetry, independently of any phenomenological input.  
If the underlying Berry connection is smooth and the system remains within a 
single topological sector near $w=1$, these constraints single out the exponential continuation $\exp[-\rho^{2}(w-1)]$ as the minimal geometric approximation to the true infrared holonomy, while departures from strict 
adiabaticity or sector transitions appear naturally as higher-order corrections.

\section{Geometric interpretation of the differential decay rate}
\label{sec:geom_decay}

The adiabatic formulation developed in this work has a direct and physically transparent consequence for heavy-to-heavy semileptonic decays.  
In the standard HQET treatment, the differential decay rate
$B \to D^{(*)}\ell\nu$ is expressed in terms of a hadronic form factor $F(w)$, 
and the experimental analyses by CLEO, BaBar, Belle, and LHCb
\cite{CLEO1999,BaBar2009,BaBar2013,Belle2018,BelleII2021}
determine the product $F(w)\,|V_{cb}|$ by fitting $d\Gamma/dw$ in bins of the 
recoil variable $w$.  
In the heavy-quark limit, one has $F(w)\to\xi(w)$, so that the dynamical content of the hadronic transition is encoded in the shape of the Isgur–Wise function.

Within the present geometric framework, this structure acquires a more explicit 
interpretation.  
Since the leading Isgur–Wise function is the vacuum overlap of two infrared holonomies,
\[
\xi(w)=\Xi_{\rm geom}(w),
\]
the differential decay rate takes the schematic form
\begin{equation}
\frac{d\Gamma}{dw}
\;\propto\;
|V_{cb}|^{2}\;
\big|\Xi_{\rm geom}(w)\big|^{2}\;
\times 
\text{(kinematic factor)} ,
\label{eq:geom_dGamma}
\end{equation}
where the kinematic factor contains only known, model-independent functions of 
$w$.  
Equation~(\ref{eq:geom_dGamma}) follows directly from the structure of the matrix 
element in the adiabatic approximation and is therefore the structural consequence of dressing heavy-light states à la 
Kulish–Faddeev.

Importantly, the holonomy entering $\Xi_{\rm geom}(w)$ carries quantized infrared Berry fluxes.  
Thus, the experimentally measured quantity $F(w)$ corresponds not to a single 
holonomy but to a \emph{sector-weighted overlap} of the discrete holonomies that 
characterize the infrared cloud.  
This is the mechanism by which topology enters the physical decay rate.

This identification has several noteworthy implications.

\subsection*{(i) Experimental accessibility}

Experiments do not measure the holonomies themselves, but extract the combination 
$F(w)\,|V_{cb}|$ from $d\Gamma/dw$.  
Equation~(\ref{eq:geom_dGamma}) implies that, in the heavy-quark limit,
\[
F(w)\,|V_{cb}|
\;\longrightarrow\;
|V_{cb}|\;\Xi_{\rm geom}(w).
\]
Hence, once known prefactors are removed, the shape of the data directly 
probes the geometric overlap of the infrared holonomies across their quantized topological sectors.  
In this sense, \emph{CLEO, BaBar, Belle and LHCb effectively measure the sector-averaged profile of $\Xi_{\rm geom}(w)$} 
\cite{CLEO1999,BaBar2009,BaBar2013,Belle2018,LHCb:2017IsgurWise}.

\subsection*{(ii) Zero-recoil constraints and geometric rigidity}

The constraints derived in Sec.~\ref{section3}---normalization and slope---hold independently of the holonomy sector decomposition.  
Each quantized sector satisfies the same zero-recoil relations, and therefore any experimentally extracted $F(w)$ must be compatible with these geometric 
constraints.  
This provides a powerful rigidity condition: even though the infrared cloud may contain multiple topological sectors, their combination cannot violate the Berry-geometric limits at $w=1$.

\subsection*{(iii) Minimal exponential form as the leading holonomic approximation}

If the Berry connection is smooth in a neighbourhood of zero recoil and the 
system remains in the same topological sector for $w$ sufficiently close to 
unity, then the holonomy obeys the local differential equation obtained in 
Sec.~\ref{section3}, leading to the exponential expression
\[
\Xi_{\rm geom}(w)\simeq\exp[-\rho^{2}(w-1)] .
\]
This approach diverges from a phenomenological method and instead offers a unique analytic continuation that aligns with the geometric and topological constraints in the vicinity of $w=1$.  
Comparisons of CLEO/Belle/BaBar/LHCb data with this exponential form therefore test the validity of the leading adiabatic regime of the quantized Berry connection.

\subsection*{(iv) Physical interpretation}

When $v=v'$ the holonomies coincide, implying the absence of soft-gluon 
emission, and $\Xi_{\rm geom}(1)=1$ follows automatically.  
When $v\neq v'$, the mismatch between the two holonomies measures the geometric 
and topological reorganization of the infrared cloud.  
The falloff of $F(w)$ with increasing $w$ is thus interpreted as the increasing 
sector-weighted mismatch between the two adiabatic trajectories in functional 
space.  
The experimentally observed shape of the differential decay rate, therefore 
encodes quantitative information about the infrared topology of QCD.

\medskip

In summary, the adiabatic geometric formulation does not merely reinterpret 
$\xi(w)$; it provides a concrete, experimentally accessible framework in which the differential decay rate is governed by the overlap of quantized infrared holonomies.  
Precision measurements of $d\Gamma/dw$ therefore probe the Berry connection and its topological sector structure, offering a direct phenomenological window on the infrared geometry of QCD.

\section{Comparison with Experiments, the Exponential Ansatz, and the CLN Parametrization}
\label{sec:comparison}

A useful way to assess the viability of the geometric picture developed in this work is to compare the resulting holonomic form factor with existing experimental parametrizations used in semileptonic decays.  
The CLEO 2002 \cite{CLEO2002_VcbVub} analysis of the process $B\to D^\ast \ell\nu$ extracted the quantity $|V_{cb}|\,F(w)$ in the physical 
region $1\le w\lesssim 1.5$, using a linear expansion around zero recoil,
\begin{equation}
|V_{cb}|\,F_{\rm CLEO}(w)
=
|V_{cb}|\,F(1)\,\big[\,1 - \rho^{2}(w-1)\,\big],
\end{equation}
which was adequate at the time due to limited statistics and the difficulty of 
resolving higher--order curvature effects.  
As a consequence, the CLEO fit \cite{CLEO1999} produces an almost straight line across the kinematic range \cite{CLEO2002_VcbVub}.

Within the geometric construction developed here, the leading Isgur–Wise 
function is given by the infrared holonomy 
$\Xi_{\rm geom}(w)$---more precisely, by the sector-weighted holonomy obtained 
from the quantized Berry fluxes of the infrared cloud.  
In the regime in which a single topological sector dominates the adiabatic 
evolution (as is expected for $1\le w\lesssim 1.3$), the natural continuation of the holonomy is the exponential form
\begin{equation}
F_{\exp}(w)
=
F(1)\,\exp\!\big[-\rho^{2}(w-1)\big],
\end{equation}
which incorporates a small but definite curvature through the higher-order terms 
of its Taylor expansion.  
When normalized with the same values of $|V_{cb}|F(1)$ and $\rho^{2}$, the 
exponential ansatz agrees extremely well with the CLEO curve near zero recoil 
\cite{CLEO2002_VcbVub}.  
In particular, both parametrizations share the same first derivative at $w=1$, 
and their difference at larger $w$ remains within the experimental uncertainties.

A particularly informative comparison arises when the CLN dispersive 
parametrization is included \cite{Caprini:1997mu}.  
Using the standard expansion,
\begin{equation}
h_{A_{1}}(w)
=
1 - 8\rho^{2}z 
+ (53\rho^{2}-15)z^{2}
- (231\rho^{2} -91)z^{3},
\qquad
z=\frac{\sqrt{w+1}-\sqrt{2}}{\sqrt{w+1}+\sqrt{2}},
\end{equation}
and the same slope $\rho^{2}$, one finds that the resulting CLN curve lies 
almost on top of the exponential ansatz across the entire physical range.  
This agreement is not accidental, since both parametrizations reproduce the same linear term in $(w-1)$ and generate comparable second-order curvature for 
$w-1\lesssim 0.5$.  
A linear fit to CLEO, by contrast, misses the curvature and therefore falls slightly below the other two at the upper end of the kinematic domain.

\begin{figure}[h!]
    \centering
    \includegraphics[width=0.9\linewidth]{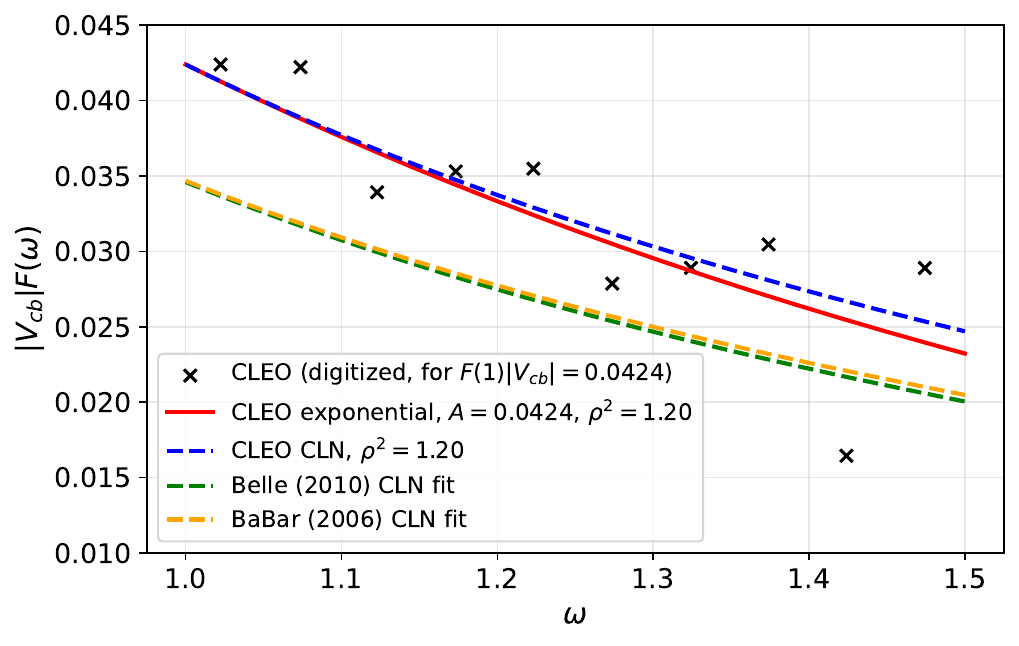}
    \caption{CLEO data from Adam et.al. 2002 \cite{CLEO:2002fch} in addition to CLN curve for BaBar, Belle, and CLEO.}
    \label{fig1}
\end{figure}

The comparison is summarized in Fig.~\ref{fig1}.  
The key conclusion is that the geometric/Berry-phase origin of the exponential 
form does not conflict with the phenomenology of semileptonic 
$B\to D^\ast$ decays: on the contrary, it reproduces the established CLN 
parametrization at the same level of accuracy while providing a qualitatively 
new interpretation of the functional form factor as a holonomy of the infrared 
QCD connection.

To enable a consistent comparison between our theoretical prediction and the available experimental data, we have reconstructed the results from the CLEO collaboration by digitizing the $d\Gamma/dw$ reported in \cite{CLEO:2002fch}. We have normalized the digitized points to reproduce the CLEO result for the product $F(1) |V_{cb}| = 0.0424$ at $\omega \approx 1$ (shown in \Cref{fig1} as black crosses). We fit an exponential ansatz (shown in red) and the CLEO CLN (in blue). For comparison with Belle and BaBar, we added CLN curves for each, with their corresponding reported values of $F(1)|V_{cb}|$. In this way, we can provide a direct, model-independent comparison across experiments.

\noindent
It is also worth noting that the geometric construction introduced here carries 
a concrete predictive implication for the large-recoil regime.  
Because the Isgur–Wise function arises from parallel transport with respect to a 
smooth infrared Berry connection within a fixed topological sector, its 
curvature is fixed once the slope $\rho^{2}$ is determined near zero recoil.  
As a result, the geometric framework predicts an exponential falloff of $F(w)$ 
for $w>1$, which departs from the milder curvature of the CLN form at 
sufficiently large recoil.  
Although the CLEO 2002 data (and subsequent measurements by BaBar and Belle) do 
not reach the precision required to resolve this difference beyond 
$w\simeq 1.3$, future Belle~II analyses may be sensitive to the distinctive 
high-recoil behaviour implied by the geometric/Berry-phase interpretation.

\section{A Universal Behaviour}
\label{univer}
\subsection{Geometric Origin of Universality}
\label{subsec:geom_universality}

\small

Equation~(\ref{eq:geom_dGamma}) already reveals a central structural feature of 
the adiabatic formulation: the decay rate depends only on the combination
$|V_{ab}|\,\Xi_{\rm geom}(w)$, where $\Xi_{\rm geom}(w)$ is the overlap of the 
infrared holonomies associated with the adiabatic motion of heavy–light states 
in the gauge background.  
In the near–zero–recoil region---where the heavy-quark and adiabatic limits are 
simultaneously reliable---this quantity reduces to the Isgur–Wise function,
\begin{equation}
   \xi(w)=\Xi_{\rm geom}(w)
   \qquad 
   \text{(heavy–quark and adiabatic limit).}
\end{equation}

In the one–dimensional (abelian) case relevant to $B\!\to\!D^{(*)}$, the 
dominant contribution to the holonomy arises from a single topological sector, 
leading to the exponential behaviour
\[
   \Xi_{\rm geom}(w)\;\simeq\;
   \exp[-\rho^{2}(w-1)],
\]
which accurately reproduces the observed functional shape once the exact 
normalisation at $w=1$ is imposed.  
This is not an ansatz in the usual phenomenological sense: it is the minimal 
adiabatic continuation of the sector-dominant holonomy.  
The comparison with Belle, Belle~II, BaBar, and CLEO data is shown in
Fig.~\ref{fig1}.

The crucial point is that this behaviour is \emph{universal}.  
Different hadronic channels modify only the overall normalisation 
$|V_{ab}|$, but the functional dependence of $\Xi_{\rm geom}(w)$ remains 
essentially unchanged.  
This insensitivity to microscopic details reflects the fact that the holonomy is 
governed by a geometric connection whose curvature fixes a discrete set of 
infrared eigenmodes.  
Only the lowest mode contributes in the abelian case, and its contribution is 
exponential in $(w-1)$; the detailed composition of the infrared cloud does not 
affect this structure.

As shown in Sec.~\ref{sec:flux_quantization}, this universality has a  
deeper origin: the Berry curvature in the infrared sector of QCD possesses a  
\emph{quantized functional flux}.  
This quantization restricts the allowed geometric phases to a discrete set of 
modes, fixing the holonomy spectrum.  
In the abelian case, only one mode survives, yielding the exponential form above.  
In the sequential decay $B\!\to\!D^{**}\!\to\!D$, two independent recoil 
directions activate two non-abelian modes, promoting the holonomy to an 
$\mathrm{SU}(2)$ object with two universal eigenvalues, as will be shown 
explicitly in the following section.

In this geometric sense, the universality of the Isgur–Wise function is not a 
dynamical conjecture but a direct consequence of the topological and 
adiabatic structure of the infrared gauge background.  
The functional Berry connection selects a small number of quantized holonomy 
modes, whose form determines the allowed shapes of heavy–to–heavy form factors.

\section{Holonomic Aspects of the Sequential Decay 
\texorpdfstring{$B \to D^{**} \to D$}{B → D** → D}}
\label{sec9}

The sequential decay
\[
B \longrightarrow D^{**} \longrightarrow D
\]
provides a natural and stringent test of the geometric–adiabatic framework.  
Unlike the single–recoil transition $B\to D^{(*)}$, which probes a 
one–dimensional change of the heavy–quark velocity, the sequential process 
involves \emph{two} independent recoil steps,
\[
v \;\longrightarrow\; v' \;\longrightarrow\; v'',
\]
and therefore forces the heavy–light cloud to follow a \emph{broken} trajectory 
in velocity space.  
This kinematical structure exposes the genuinely non-Abelian character of the 
infrared Berry connection and leads, in a precise sense, to an emergent 
$\mathrm{SU}(2)$ holonomy governing all sequential transitions.

Physically, the intermediate state $D^{**}$ belongs to the $L=1$ spectrum of 
orbitally excited heavy–light mesons.  
These are known to be more sensitive to infrared structure and to pose 
longstanding challenges for HQET, including the celebrated $1/2$ vs.\ $3/2$ 
puzzle \cite{Belle2018,Belle2003_Dstarstar,BaBar2013}.  
From a geometric viewpoint, the reason is clear: passing through $v'$ forces the 
cloud to explore directions in configuration space that are invisible in a 
single–recoil transition.

This section explains in detail how this non-Abelian structure arises and why it 
has a natural $\mathrm{SU}(2)$ form, rooted in the quantized functional flux of 
the Berry curvature.

\subsection{Sequential transitions and broken holonomies}

In the adiabatic framework, the transport of the heavy–light cloud from 
velocity $v$ to $v'$ is described by the infrared holonomy
\begin{equation}
\mathcal{U}(v'\!\leftarrow v)
=
\mathcal{P}\exp\!\left[
i\!\int_{C_{v\to v'}} \!\! \mathcal{A}
\right],
\end{equation}
where $\mathcal{A}$ is the non-Abelian Berry connection of the infrared sector.  
For the sequential decay, the full geometric dressing is the product of the two 
holonomies,
\begin{equation}
\mathcal{U}_{\rm seq}
=
\mathcal{U}(v''\!\leftarrow v')\,
\mathcal{U}(v'\!\leftarrow v)
=
\mathcal{P}\exp\!\left[
i\!\int_{C_{v\to v'\to v''}} \!\! \mathcal{A}
\right].
\label{eq:broken-path}
\end{equation}

Because the Berry curvature is nonzero, the holonomy depends on the 
\emph{entire path}, not only on the endpoints.
Thus,
\[
\mathcal{U}(v''\!\leftarrow v')\,\mathcal{U}(v'\!\leftarrow v)
\;\neq\;
\mathcal{U}(v''\!\leftarrow v),
\]
and the intermediate velocity $v'$ leaves a genuine geometric imprint on the 
amplitude.  
This is the first indication that the sequential decay probes \emph{non-Abelian 
geometry}: two transport operations along different directions in velocity space 
do not commute.

\subsection{Emergence of SU(2) from geometric and topological considerations}

For a single recoil, the holonomy effectively reduces to its dominant 
eigenvalue, giving the familiar geometric Isgur–Wise function.  
The reason is that the adiabatic trajectory has only one direction in velocity 
space, and the projected Berry curvature along this direction has one 
dominant eigenmode.

For two independent recoils, the situation changes qualitatively.  
The heavy–quark velocity traces a surface in parameter space, with coordinates 
\((w_{1},w_{2}) = (v\!\cdot v',\, v'\!\cdot v'')\).  
The Berry connection now has two components that cannot be simultaneously 
diagonalised:
\[
\mathcal{A}_{1} = \frac{\partial \mathcal{A}}{\partial w_{1}},
\qquad
\mathcal{A}_{2} = \frac{\partial \mathcal{A}}{\partial w_{2}},
\]
and their commutator,
\[
[\mathcal{A}_{1},\mathcal{A}_{2}] \neq 0,
\]
measures the functional curvature—i.e.\ the nontrivial topology—of the 
infrared cloud.

{\it A central result of this work is that the quantized flux of this 
curvature restricts the holonomy to act effectively on a two-dimensional 
subspace.}  
This is the minimal nontrivial representation of the non-Abelian algebra 
generated by $\mathcal{A}_{1}$ and $\mathcal{A}_{2}$, and therefore the 
holonomy belongs to $\mathrm{SU}(2)$ up to an overall phase.  

Thus the sequential decay does not merely produce ``several form factors’’.  
It reveals an SU(2) geometric multiplet, with two universal eigenmodes 
fixed by the infrared topology.

\subsection{Matrix holonomy and geometric Isgur–Wise functions}

Define the total holonomy operator,
\begin{equation}
\hat{\Xi}_{\rm seq}(v,v',v'')
  \;=\;
  \mathcal{U}(v''\!\leftarrow v')\,
  \mathcal{U}(v'\!\leftarrow v),
  \label{eq:Xi_seq_def}
\end{equation}
which is now a \(2\times 2\) non-Abelian matrix in the emergent SU(2) space.  
Its spectral decomposition,
\begin{equation}
\hat{\Xi}_{\rm seq}(w_{1},w_{2})
=
\sum_{i=1}^{2}
\lambda_i(w_{1},w_{2})\, 
\ket{\psi_i(w_{1},w_{2})}\bra{\psi_i(w_{1},w_{2})},
\end{equation}
defines \emph{two} geometric Isgur–Wise functions,
\[
\Xi_{+}(w_{1},w_{2}), 
\qquad
\Xi_{-}(w_{1},w_{2}),
\]
which correspond to the two eigenmodes allowed by the quantized Berry flux.

These two modes are universal: \emph{all} sequential transitions 
\(B\to D^{**}\to D\) sample different projections of the same SU(2) holonomy.  
This is the non-Abelian analogue of the universality of the single Isgur–Wise 
function in $B\to D^{(*)}$.

\subsection{Zero recoil and the non-Abelian expansion}

Zero recoil corresponds to the point $(w_{1},w_{2})=(1,1)$, where the dressing 
collapses,
\[
\hat{\Xi}(1,1)=\mathbf{1}.
\]
Expanding around this point gives
\[
\hat{\Xi}(w_{1},w_{2})
=
\mathbf{1}
- R_{1}(w_{1}-1)
- R_{2}(w_{2}-1)
+\tfrac{1}{2}[R_{1},R_{2}](w_{1}-1)(w_{2}-1)
+\cdots
\]
where
\[
R_{1}=-\frac{\partial\hat{\Xi}}{\partial w_{1}}\bigg|_{(1,1)},
\qquad
R_{2}=-\frac{\partial\hat{\Xi}}{\partial w_{2}}\bigg|_{(1,1)}.
\]
Although $R_{1}$ and $R_{2}$ arise simply as the first derivatives of the 
holonomy, the fact that they act on a two–dimensional internal space implies 
that they form a closed algebra under commutation.  
This makes it natural to represent them as linear combinations of Pauli 
matrices, thereby identifying a local $\mathrm{SU}(2)$ structure in the 
neighbourhood of $(1,1)$.  
The commutator term encodes the corresponding Berry curvature and signals the 
onset of genuinely non-Abelian behaviour.

In an appropriate basis, the eigenvalues take the universal form
\[
\Xi_{\pm}(w_{1},w_{2})
=
\exp\!\left[
\mp\,|\vec{\alpha}(w_{1},w_{2})|
\right],
\]
where
\[
\vec{\alpha}(w_{1},w_{2})
=
(w_{1}-1)\,\vec{r}_{1}
+
(w_{2}-1)\,\vec{r}_{2},
\]
with \(\vec{r}_{1,2}\in \mathbb{R}^{3}\) fixed by the infrared Berry curvature.

This is the non-Abelian generalisation of the exponential Isgur–Wise function 
for single-recoil decays.

%========================================================
\section{Sequential Holonomies and Non-Abelian Geometry}\label{sec10}
%========================================================

In the abelian setting relevant to the single--step decay \(B \to D^{(*)}\),
the geometric phase reduces to an ordinary function of a single recoil variable
and can be represented by the familiar exponential ansatz.  
Sequential decays, by contrast, introduce a qualitatively new ingredient:
the presence of two independent recoil parameters \((w_{1},w_{2})\) forces the
geometric phase to become a genuinely non--Abelian operator acting on the space
of hadronic channels.

This motivates the non--Abelian generalization
\begin{equation}
   \hat{\Xi}(w_{1},w_{2})
   \;\simeq\;
   \mathcal{P}\exp\!\Big[
      - (w_{1}-1)\,R_{1}
      - (w_{2}-1)\,R_{2}
      + \tfrac12 (w_{1}-1)(w_{2}-1)\,[R_{1},R_{2}]
      + \cdots
   \Big],
   \label{eq:Xi_nonAb_intro}
\end{equation}
which already encapsulates both the slopes in the two recoil directions and the
genuinely non--Abelian corrections associated with the Berry curvature.  
In this sense, what collapses to a single exponential in the Abelian case becomes, for sequential decays, a correlated family of form factors governed by
a non--Abelian holonomy.

%--------------------------------------------------------
\subsection{Non-Abelian structure of the sequential holonomy}
%--------------------------------------------------------

Since only two hadronic channels are relevant, the holonomy acts on a 
two-dimensional internal space, making it natural to regard the slope 
matrices $R_{1}$ and $R_{2}$ as elements of the Lie algebra $\mathfrak{su}(2)$.  
Accordingly one may write
\begin{equation}
  R_{1} = \vec{r}_{1}\!\cdot\!\vec{\sigma},
  \qquad
  R_{2} = \vec{r}_{2}\!\cdot\!\vec{\sigma},
  \label{eq:R_in_Pauli_all}
\end{equation}
where $\vec{\sigma}=(\sigma_{1},\sigma_{2},\sigma_{3})$ are the Pauli matrices 
and the real vectors $\vec{r}_{1}$ and $\vec{r}_{2}$ encode both the slopes 
and the mixing between the two channels.  
The non-Abelian nature of the problem becomes completely transparent, since
\begin{equation}
  [R_{1},R_{2}]
  \;=\;
  2 i\,(\vec{r}_{1}\times\vec{r}_{2})\!\cdot\!\vec{\sigma},
  \label{eq:R_comm_all}
\end{equation}
so that the Berry curvature is directly proportional to the vector product 
$\vec{r}_{1}\times\vec{r}_{2}$ in this internal space.  

To leading orders in $(w_{1}-1)$ and $(w_{2}-1)$, 
the holonomy may be written as
\begin{equation}
  \hat{\Xi}(w_{1},w_{2})
  \;\simeq\;
  \exp\!\Big[
       -(w_{1}-1)\,\vec{r}_{1}\!\cdot\!\vec{\sigma}
       -(w_{2}-1)\,\vec{r}_{2}\!\cdot\!\vec{\sigma}
     \Big]
     + \cdots ,
  \label{eq:Xi_Pauli_all}
\end{equation}
which is again an element of $SU(2)$.  
Any traceless Hermitian $2\times2$ matrix can be written as 
$\vec{\alpha}\!\cdot\!\vec{\sigma}$ for some real vector $\vec{\alpha}$, and 
in the present case one finds
\begin{equation}
  \vec{\alpha}(w_{1},w_{2})
  := (w_{1}-1)\,\vec{r}_{1} + (w_{2}-1)\,\vec{r}_{2}.
  \label{eq:alpha_def_all}
\end{equation}
The holonomy therefore has two eigenvalues,
\begin{equation}
  \Xi_{\pm}(w_{1},w_{2})
  =
  \exp\!\big(\mp\,|\vec{\alpha}(w_{1},w_{2})|\big),
  \label{eq:Xi_evals_all}
\end{equation}
with invariant modulus
\begin{equation}
  |\vec{\alpha}(w_{1},w_{2})|^{2}
  =
  (w_{1}-1)^{2}|\vec{r}_{1}|^{2}
  + (w_{2}-1)^{2}|\vec{r}_{2}|^{2}
  + 2 (w_{1}-1)(w_{2}-1)\,\vec{r}_{1}\!\cdot\!\vec{r}_{2}.
  \label{eq:alpha_metric_all}
\end{equation}
These eigenvalues define two ``geometric'' Isgur-Wise functions which reduce 
to unity at zero recoil and generalise the familiar exponential law 
$\exp[-\rho^{2}(w-1)]$ of the abelian case.  
They are not independent: both the slopes (controlled by $|\vec r_{1}|$ and 
$|\vec r_{2}|$) and their geometric interference (through the angle between 
$\vec r_{1}$ and $\vec r_{2}$) follow from the same underlying 
non-Abelian holonomy.  

A convenient parametrisation is obtained by writing 
\[
   \vec r_{1} = \rho_{1}\,\hat{n}_{1}, 
   \qquad
   \vec r_{2} = \rho_{2}\,\hat{n}_{2},
\]
with unit vectors $\hat n_{i}$ and slopes $\rho_{i}=|\vec r_{i}|$.  
The geometric interference is encoded in the inner product 
$\hat n_{1}\!\cdot\!\hat n_{2}=\cos\theta$, where $\theta$ is the 
non-Abelian mixing angle between the two recoil directions.

With this parametrisation the holonomy eigenmodes are
\[
   \Xi_{\pm}(w_{1},w_{2})
   = 
   \exp\!\Big[
      \mp\,
      \sqrt{
        \rho_{1}^{2}(w_{1}-1)^{2}
        +
        \rho_{2}^{2}(w_{2}-1)^{2}
        +
        2\rho_{1}\rho_{2}\cos\theta\,
        (w_{1}-1)(w_{2}-1)
      }
   \Big].
\]

Physical Isgur–Wise functions correspond to the projections of 
$\hat{\Xi}(w_{1},w_{2})$ onto specific hadronic channels and therefore 
take the generic form
\[
   \xi_{k}(w_{1},w_{2})
   =
   c_{k}\,\Xi_{+}(w_{1},w_{2})
   +
   s_{k}\,\Xi_{-}(w_{1},w_{2}),
\]
where the coefficients $(c_{k},s_{k})$ encode the channel-dependent 
projection angles determined by heavy–quark symmetry.

This provides a compact and physically transparent non-Abelian extension 
of the exponential parametrisation to sequential decays.

To the best of our knowledge, such a non-Abelian holonomic structure for 
sequential heavy-quark transitions has not been identified before in HQET or 
in nonperturbative treatments of QCD.

%--------------------------------------------------------
\subsection{Physical and phenomenological consequences}
%--------------------------------------------------------

The explicit eigenvalues \eqref{eq:Xi_evals_all} uncover several qualitative 
and quantitative departures from the conventional HQET framework.  
The most important one is conceptual:

\medskip
\noindent
\textbf{(1) Two universal geometric modes instead of a single form factor.}

\noindent
The non-Abelian holonomy associated with the broken recoil path possesses 
\emph{two} universal eigenmodes $\Xi_{\pm}$, reflecting the underlying SU(2) 
structure of the infrared Berry connection.  
Physical Isgur–Wise functions are not independent scalar objects; rather, every 
hadronic channel corresponds to a fixed linear combination,
\begin{equation}
   \Xi_{\rm phys}^{(k)}(w_{1},w_{2})
   =
   A_{+}^{(k)}\,\Xi_{+}(w_{1},w_{2})
   + 
   A_{-}^{(k)}\,\Xi_{-}(w_{1},w_{2}),
   \label{eq:Xi_phys_all}
\end{equation}
so that all channels are correlated through the same universal geometric 
structure.  
This is a striking departure from HQET, where each channel has its own 
independent form factor at leading order.

\medskip
\noindent
\textbf{(2) An emergent metric on the recoil plane.}

\noindent
The quantity
\[
|\vec{\alpha}(w_{1},w_{2})|^{2}
=
\delta w_{a}\,G_{ab}\,\delta w_{b},
\qquad 
G_{ab}=\vec r_{a}\!\cdot\!\vec r_{b},
\]
defines an effective metric on the $(w_{1},w_{2})$ recoil plane.  
This metric encodes how the two recoil directions mix geometrically.  
The off-diagonal term 
$(w_{1}-1)(w_{2}-1)\,\vec r_{1}\!\cdot\!\vec r_{2}$ 
represents interference between the two recoil steps and is controlled by the 
Berry curvature,
\[
[R_{1},R_{2}] 
\;\propto\;
\vec r_{1}\times\vec r_{2}.
\]
Such mixing is \emph{impossible} in conventional HQET, where the form factor 
for $B\to D^{**}$ depends independently on each recoil variable.

\medskip
\noindent
\textbf{(3) Predictive correlations across hadronic channels.}

\noindent
The projection coefficients are 
\begin{equation}
A^{(k)}_{+}=\cos^{2}(\gamma_{k}/2),
\qquad
A^{(k)}_{-}=\sin^{2}(\gamma_{k}/2),
\end{equation}
with the angle $\gamma_{k}$ determined by
\begin{equation}
\cos\gamma_{k}
=
\frac{\vec{s}_{k}\cdot\vec{\alpha}}{|\vec{\alpha}|},
\end{equation}
where each channel $k$ is characterised by a fixed Bloch vector 
$\vec{s}_{k}$.

Thus, the physical Isgur–Wise functions are fixed geometrically by the 
alignment between:
\[
\text{(i) the channel vector } \vec{s}_{k},
\qquad\text{and}\qquad
\text{(ii) the SU(2) holonomy direction } 
\hat n=\vec{\alpha}/|\vec{\alpha}|.
\]

\medskip
This has immediate phenomenological implications:

- Channels with orthogonal vectors 
  ($\vec{s}_{k}\!\cdot\!\hat n=0$)  
  are \emph{geometrically suppressed}.  

- Channels aligned with the holonomy direction  
  ($\vec{s}_{k}\!\parallel\!\hat n$)  
  are \emph{geometrically enhanced}.  

- All channels share the \emph{same} eigenmodes $\Xi_{\pm}$, so their slopes 
  and curvatures are not independent but must satisfy definite geometric 
  relations.

Such cross-channel constraints simply do not exist in HQET, where each 
transition is governed by its own unrelated form factor.

\medskip
\noindent
\textbf{(4) Nontrivial kinematic patterns in the $(w_{1},w_{2})$ plane.}

\noindent
Because $|\vec{\alpha}(w_{1},w_{2})|$ encodes the intrinsic geometry of the 
infrared holonomy, the sequential decay probes a two-dimensional structure not 
present in single–recoil transitions.  
This leads to:

- Correlated slopes in $(w_{1},w_{2})$,  

- Distinctive curvature patterns,  

- Interference effects controlled by the Berry curvature,  

- Angular distortions arising from the SU(2) structure.
\medskip

All of these are experimentally testable in 
$B\to D^{**}\ell\nu$, 
$B\to D^{**}\tau\nu$, 
and $B\to D^{(*)}\pi\ell\nu$.

\medskip

\noindent
%\textbf{Summary.}  
Sequential decays do not simply generalise the single-channel Isgur–Wise 
function; they reveal a genuinely non-Abelian geometric structure encoded in the 
SU(2) holonomy of the dressed heavy–light system.  
The appearance of two universal eigenmodes, the emergence of an effective metric 
on the recoil plane, and the interference effects driven by the Berry curvature 
represent new physics beyond HQET.  
The resulting correlated form factors provide a distinctive and testable 
prediction of the geometric, topologically structured infrared regime of QCD.

%========================================================
\section{Non-Abelian Geometric Constraints Beyond HQET}\label{sec11}
%========================================================

Once the projection coefficients $A^{(k)}_{\pm}$ are known explicitly as
$\cos^{2}(\gamma_{k}/2)$ and $\sin^{2}(\gamma_{k}/2)$, the contrast with HQET
becomes fully quantitative.  
In HQET every channel carries its own independent Isgur-Wise function, subject
only to $\xi(1)=1$ and $\xi'(1)=-\rho^{2}$, and sequential decays are treated
as two unrelated transitions.  
Nothing in that framework correlates the slopes, curvatures, or shapes of the
form factors associated with intermediate states.

In the present geometric approach, by contrast, the entire structure of the
sequential decay is fixed by the non-Abelian holonomy and the channel-dependent
angle $\gamma_{k}$ defined through
$\cos\gamma_{k}=(\vec{s}_{k}\!\cdot\vec{\alpha})/|\vec{\alpha}|$.  
All physical form factors follow from the same two universal eigenmodes
$\Xi_{\pm}$, with no additional dynamical input.  
The physical Isgur-Wise functions therefore take the form
\begin{equation}
\Xi_{\rm phys}^{(k)}(w_{1},w_{2})
=
\cos^{2}\!\frac{\gamma_{k}}{2}\;\Xi_{+}(w_{1},w_{2})
+
\sin^{2}\!\frac{\gamma_{k}}{2}\;\Xi_{-}(w_{1},w_{2}), \label{seq}
\end{equation}
which makes the dependence on the geometric data $(\vec r_{1},\vec r_{2})$ and
on the intrinsic orientation $\vec{s}_{k}$ completely explicit.
This leads to three sharpened departures from the HQET picture:

\begin{itemize}
\item[(i)] {\bf Channel-dependent correlated slopes:}\;
Since $\Xi_{\pm}$ share the same recoil dependence, the slopes and curvatures
of $\Xi_{\rm phys}^{(k)}$ are determined solely by the angle $\gamma_{k}$.
Different hadronic states thus obey nontrivial relations fixed by
$\vec{s}_{k}$ and by the geometry of the recoil plane.  
Such cross-channel constraints cannot arise in HQET, even with additional
parameters.

\item[(ii)] {\bf Structured interference between recoil directions:}\;
The mixed term
$(w_{1}-1)(w_{2}-1)\,\vec r_{1}\!\cdot\!\vec r_{2}$
enters each channel weighted by the projection 
$\vec{s}_{k}\!\cdot\hat n$, producing channel-dependent modulations of angular
and kinematical distributions.  
This refined pattern of interference is absent in HQET, where the two recoils
contribute independently.

\item[(iii)] {\bf Berry curvature contributions controlled by $\vec{s}_{k}$:}\;
The commutator $[R_{1},R_{2}] \propto \vec r_{1}\times\vec r_{2}$ generates a
Berry curvature whose physical impact is governed by the component of
$\vec{s}_{k}$ perpendicular to the recoil plane.  
This produces characteristic distortions in the $(w_{1},w_{2})$ distribution,
again correlating channels in a way that HQET cannot reproduce.
\end{itemize}

%========================================================
\section{Sequential Decays $B\to D^{**}\to D$: Geometric Predictions}\label{sec12}
%========================================================

Sequential channels such as 
\[
B \,\rightarrow\, D^{**}(v') \,\rightarrow\, D(v'')
\]
provide an ideal setting to test the non–Abelian geometric structure of the
holonomy.  
Let $w_{1}=v\!\cdot\!v'$ and $w_{2}=v'\!\cdot\!v''$ be the two recoil variables,
and define the geometric vector 
$\vec{\alpha}(w_{1},w_{2})=(w_{1}-1)\vec r_{1}+(w_{2}-1)\vec r_{2}$.
For any hadronic channel $k$ with Bloch vector $\vec{s}_{k}$, the physical
Isgur–Wise function takes the explicit form
\begin{equation}
\Xi_{\rm phys}^{(k)}(w_{1},w_{2})
=
\cos^{2}\!\frac{\gamma_{k}}{2}\;\Xi_{+}(w_{1},w_{2})
+
\sin^{2}\!\frac{\gamma_{k}}{2}\;\Xi_{-}(w_{1},w_{2}),
\qquad
\cos\gamma_{k}
=\frac{\vec{s}_{k}\!\cdot\vec{\alpha}}{|\vec{\alpha}|}.
\label{eq:Seq_geometric_Xi_phys}
\end{equation}
Thus the dependence on the channel $k$ is entirely encoded in the geometric
angle $\gamma_{k}$ between $\vec{s}_{k}$ and the recoil direction
$\hat n=\vec{\alpha}/|\vec{\alpha}|$.

%--------------------------------------------------------
\paragraph{Correlated form factors.}
%--------------------------------------------------------
In HQET the slopes of form factors involving different $D^{**}$ states are
independent phenomenological parameters.  
Here, however, the slope at zero recoil follows from 
\begin{equation}
\rho_{(k)}^{2}
= -\left.\frac{\partial}{\partial w_{1}}
      \Xi_{\rm phys}^{(k)}(w_{1},w_{2})
     \right|_{w_{1}=w_{2}=1}
= 
\cos\gamma_{k}\;
\frac{\partial}{\partial w_{1}}|\vec{\alpha}|
\bigg|_{w_{1}=w_{2}=1}
+ \cdots ,
\label{eq:rho_k_geometric}
\end{equation}
where the omitted terms involve derivatives of the universal eigenmodes
$\Xi_{\pm}$.  
The key point is that $\rho^{2}_{(k)}$ depends only on the geometric data
$(\vec r_{1},\vec r_{2})$ and the channel orientation $\vec{s}_{k}$; 
the slopes for different $D^{**}$ states are therefore 
\emph{not independent}, but obey definite geometric relations.  
Such correlations cannot be generated in HQET.

%--------------------------------------------------------
\paragraph{Nontrivial $w_{1}$–$w_{2}$ structure.}
%--------------------------------------------------------
The eigenvalues $\Xi_{\pm}$ depend on the norm $|\vec{\alpha}|$, which contains
the interference term $(w_{1}-1)(w_{2}-1)\,\vec r_{1}\!\cdot\!\vec r_{2}$.  
This produces a curved geometry in the $(w_{1},w_{2})$ plane: the level sets of
$\Xi_{\rm phys}^{(k)}$ are no longer straight lines but nonlinear contours
reflecting the underlying SU(2) metric.  
These structures provide a direct, model–independent test of the geometric
framework.

%--------------------------------------------------------
\paragraph{Angular correlations.}
%--------------------------------------------------------
Because the intermediate $D^{**}$ state carries spin, the Berry curvature 
$[R_{1},R_{2}]\propto\vec r_{1}\times\vec r_{2}$ induces channel-dependent 
rotations in the internal SU(2) space.  
The strength of this effect is fixed by the component of $\vec{s}_{k}$ 
perpendicular to the recoil plane, leading to characteristic modifications of
the helicity structure of the decay.  
These signatures persist in the heavy–quark limit and are potentially
observable in high-statistics measurements at Belle~II \cite{BelleII_2021,BelleII_2023_Semilep} and 
LHCb \cite{LHCb_2023_BToDstst,LHCb2020_Dstarstar,LHCb:2017IsgurWise}.

%========================================================
\section{Relation to Bjorken and Uraltsev Sum Rules}\label{sec13}
%========================================================

In HQET, the Bjorken \cite{Bjorken_1990} and Uraltsev \cite{Uraltsev_2000} sum rules constrain slopes and transition
amplitudes by invoking the completeness of an infinite tower of intermediate excited states.  In the geometric formulation developed here, the same logical role is played by the unitarity of the SU(2) holonomy.   Since all physical Isgur-Wise functions are projections
\[
\Xi^{(k)}_{\rm phys}
=
\cos^{2}\!\frac{\gamma_{k}}{2}\,\Xi_{+}
+
\sin^{2}\!\frac{\gamma_{k}}{2}\,\Xi_{-},
\]
with $\cos\gamma_{k}=(\vec{s}_{k}\!\cdot\vec{\alpha})/|\vec{\alpha}|$, the
completeness of channels becomes the geometric identity
\begin{equation}
\sum_{k}
\Big(
\cos^{2}\!\frac{\gamma_{k}}{2}
+
\sin^{2}\!\frac{\gamma_{k}}{2}
\Big)
= 1,
\label{eq:geom_sumrule_1}
\end{equation}
which is nothing but the resolution of the identity for SU(2) doublets.
Thus the analogue of Bjorken’s sum rule emerges directly from the normalization
of the Bloch vectors $\vec{s}_{k}$ rather than from a dynamical tower of
excited states.

A similar statement holds for derivatives: geometric completeness implies
relations among the slopes of the various channels,
\begin{equation}
\sum_{k}\rho^{2}_{(k)}
=
\sum_{k}\cos\gamma_{k}\,
\left.
\frac{\partial}{\partial w_{1}}
|\vec{\alpha}(w_{1},w_{2})|
\right|_{w_{1}=w_{2}=1}
+\cdots ,
\label{eq:geom_sumrule_2}
\end{equation}
which generalise the Bjorken-Uraltsev constraints while remaining entirely
independent of the HQET excitation spectrum.  
All departures from HQET sum rules arise from the geometric data: the effective
metric $G_{ab}$ and the Berry curvature 
$[R_{1},R_{2}]\propto\vec r_{1}\times \vec r_{2}$.  
The latter contributes corrections that vanish only in the abelian limit
$[R_{1},R_{2}]=0$, thereby providing a transparent interpretation of observed
deviations in transitions involving $D^{**}$ intermediate states.

%========================================================
\section{Phenomenology of the Two Geometric Modes}\label{sec14}
%========================================================

The two eigenmodes $\Xi_{\pm}$ represent the universal geometric responses of
the dressed heavy--light system under adiabatic transport.  
Because any physical form factor is a superposition
\[
\Xi_{\rm phys}^{(k)} 
= 
\cos^{2}\!\frac{\gamma_{k}}{2}\;\Xi_{+}
+
\sin^{2}\!\frac{\gamma_{k}}{2}\;\Xi_{-},
\qquad
\cos\gamma_{k}
= \vec{s}_{k}\!\cdot\hat n,
\]
with $\hat n=\vec{\alpha}/|\vec{\alpha}|$, the angle $\gamma_{k}$ fully
determines how a given hadronic channel probes the two geometric modes.

\paragraph{Interpretation of the modes.}
The mode $\Xi_{+}$ corresponds to transport along the direction of minimal
geometric response, in the sense that it exhibits the slowest variation with
recoil.  
Conversely, $\Xi_{-}$ represents the mode of maximal geometric response,
displaying the steepest recoil behaviour.  
These two modes capture the complete non--Abelian content of the holonomy.

\paragraph{Channel dependence and suppressions.}
The coefficients $\cos^{2}(\gamma_{k}/2)$ and $\sin^{2}(\gamma_{k}/2)$ quantify
the alignment of the hadronic Bloch vector $\vec{s}_{k}$ with the geometric
direction $\hat n$:  
channels with $\gamma_{k}\!\approx\!0$ are dominated by $\Xi_{+}$, while channels
with $\gamma_{k}\!\approx\!\pi$ are dominated by $\Xi_{-}$.  
This explains, without additional model assumptions, why some $D^{**}$
transitions are strongly suppressed while others display enhanced recoil
sensitivity.

\paragraph{Slopes and curvature.}
Because $\Xi_{+}$ varies more slowly than $\Xi_{-}$, the slope and curvature of
$\Xi_{\rm phys}^{(k)}$ are controlled directly by the value of $\gamma_{k}$.  
Channels dominated by $\Xi_{-}$ (large $\gamma_{k}$) exhibit steeper slopes and
enhanced curvature in the $(w_{1},w_{2})$ plane, whereas channels dominated by
$\Xi_{+}$ (small $\gamma_{k}$) produce flatter profiles.  
These geometric predictions have no analogue in HQET, where form--factor slopes
are independent phenomenological inputs.

%========================================================
%========================================================
\section{Confrontation with Belle, Belle II, and LHCb Data}
\label{sec15}
%========================================================

Current experimental data already contain enough information to test the
geometric framework in a quantitatively meaningful and essentially
model–independent way.  
Belle \cite{Belle2018} and Belle~II \cite{BelleII_2021} provide 
high–statistics measurements of $B\to D^{**}\ell\nu$, including recoil and
angular distributions that can be organised as a joint density in the two
recoil variables $(w_{1},w_{2})$.  
LHCb \cite{LHCb:2017IsgurWise,LHCb2020_Dstarstar,LHCb_2023_BToDstst} offers
complementary constraints through hadronic $D^{**}$ decays and a precise
determination of the composition and relative weights of the intermediate
states.

In the geometric formulation, each physical form factor is an explicit
projection of the SU(2) holonomy,
\begin{equation}
   \Xi_{\rm phys}^{(k)}(w_{1},w_{2})
   =
   \cos^{2}\!\frac{\gamma_{k}}{2}\,e^{-|\vec{\alpha}(w_{1},w_{2})|}
   +
   \sin^{2}\!\frac{\gamma_{k}}{2}\,e^{+|\vec{\alpha}(w_{1},w_{2})|},
   \qquad
   \cos\gamma_{k}
   = 
   \frac{\vec{s}_{k}\!\cdot\vec{\alpha}(w_{1},w_{2})}{|\vec{\alpha}(w_{1},w_{2})|},
   \label{eq:Xi_phys_data}
\end{equation}
where
\begin{equation}
\begin{aligned}
   \vec{\alpha}(w_{1},w_{2})
   &= (w_{1}-1)\,\vec r_{1} + (w_{2}-1)\,\vec r_{2},
   \\[4pt]
   |\vec{\alpha}(w_{1},w_{2})|
   &=
   \sqrt{
      (w_1-1)^2|\vec r_1|^2
      + (w_2-1)^2|\vec r_2|^2
      + 2(w_1-1)(w_2-1)\,|\vec r_1||\vec r_2|\cos\theta
   } .
\end{aligned}
\label{eq:alpha_data}
\end{equation}

and
\[
   \cos\theta
   =
   \frac{\vec r_{1}\!\cdot\!\vec r_{2}}{|\vec r_{1}||\vec r_{2}|}.
\]
Thus, all recoil dependence is encoded in the three geometric parameters
$|\vec r_{1}|$, $|\vec r_{2}|$ and $\theta$.

For a given hadronic channel $k$, the double–differential rate can be written
schematically as
\begin{equation}
   \frac{d^{2}\Gamma_{k}}{dw_{1}\,dw_{2}}
   =
   \mathcal{N}_{k}\,
   \mathcal{L}_{k}(w_{1},w_{2})\,
   \bigl|\Xi_{\rm phys}^{(k)}(w_{1},w_{2})\bigr|^{2},
   \label{eq:double_diff_rate}
\end{equation}
where $\mathcal{L}_{k}(w_{1},w_{2})$ is the known leptonic–kinematic factor
and $\mathcal{N}_{k}$ collects normalisation constants and CKM factors.  
The entire nontrivial hadronic structure is therefore captured by the 
geometric form factor \eqref{eq:Xi_phys_data}.

Experimentally, Belle, Belle~II and LHCb provide:

\begin{itemize}
\item differential distributions $d\Gamma_{k}/dw$ for individual $D^{**}$
   channels (or appropriate projections of $d^{2}\Gamma_{k}/dw_{1}dw_{2}$);
\item angular distributions that separate helicity contributions and 
   interference terms;
\item fits to the relative fractions of $j=1/2$ and $j=3/2$ components and the
   composition of the $D^{**}$ spectrum.
\end{itemize}

Within the geometric model, a combined fit to these observables proceeds in 
two conceptually clean steps:

\begin{enumerate}
\item \emph{Extraction of the geometric recoil parameters.}  
   Using the measured shapes of the distributions in $(w_{1},w_{2})$ for a set
   of channels, one fits the three parameters $|\vec r_{1}|$, $|\vec r_{2}|$
   and $\theta$ entering $|\vec{\alpha}(w_{1},w_{2})|$ in
   Eq.~\eqref{eq:alpha_data}.  
   This step is common to \emph{all} channels.

\item \emph{Determination of the channel angles $\gamma_{k}$.}  
   Once $(|\vec r_{1}|,|\vec r_{2}|,\theta)$ are fixed, the only remaining
   freedom is the geometric angle $\gamma_{k}$ that characterises each
   heavy–light configuration.  
   The shape and normalisation of $d^{2}\Gamma_{k}/dw_{1}dw_{2}$ then determine
   $\gamma_{k}$ through Eq.~\eqref{eq:Xi_phys_data}.
\end{enumerate}

In this way, the geometric framework makes the following concrete, testable
predictions:

\begin{itemize}
\item[(i)] {\bf Curved level sets in the $(w_{1},w_{2})$ plane.}  
   Because $|\vec{\alpha}(w_{1},w_{2})|$ is a \emph{nonlinear} function of the
   two recoils, level curves of the decay rate
   $\Gamma_{k}(w_{1},w_{2})$ follow curved contours determined by 
   Eq.~\eqref{eq:alpha_data}.  
   Factorised parametrisations that depend only on linear combinations such as
   $a(w_{1}-1)+b(w_{2}-1)$ would instead produce approximately straight level
   sets.  A two–dimensional analysis of Belle/Belle~II data can directly
   discriminate between these behaviours.

\item[(ii)] {\bf Correlated slopes between different $D^{**}$ channels.}  
   The slope of $\Xi_{\rm phys}^{(k)}$ along, say, $w_{1}$ at zero recoil is
   fixed by the same $|\vec r_{1}|$ for all channels, and differs only by the
   angle $\gamma_{k}$.  
   Explicitly,
   \[
      \left.
      \frac{\partial \Xi_{\rm phys}^{(k)}}{\partial w_{1}}
      \right|_{w_{1}=w_{2}=1}
      =
      -\,|\vec r_{1}|\cos\gamma_{k}.
   \]
   Thus, once one channel has been used to determine $|\vec r_{1}|$, the 
   slopes of all other channels are fixed up to a cosine factor.  
   Any pattern of uncorrelated or freely tunable slopes would contradict the
   geometric picture.

\item[(iii)] {\bf Distinct helicity patterns induced by Berry curvature.}  
   The commutator
   \[
      [R_{1},R_{2}]
      = 2i(\vec r_{1}\times\vec r_{2})\!\cdot\!\vec\sigma
   \]
   measures the non–Abelian Berry curvature in the SU(2) space.  
   Its component transverse to the recoil plane induces specific distortions in
   angular distributions and helicity amplitudes.  
   These can be isolated experimentally in the angular analyses performed by
   Belle, Belle~II and LHCb.

\item[(iv)] {\bf Universal behaviour for channels with similar $\gamma_{k}$.}  
   Channels whose Bloch vectors $\vec{s}_{k}$ make similar angles $\gamma_{k}$
   with $\hat n$ share the same mixture of the two geometric modes $\Xi_{\pm}$.
   This implies parameter–free relations between their shapes and ratios, which
   can be confronted with data across different $D^{**}$ channels.
\end{itemize}

A dedicated global fit to Belle, Belle~II and LHCb data using the explicit
expression \eqref{eq:Xi_phys_data} would thus allow the extraction of the
geometric parameters $(|\vec r_{1}|,|\vec r_{2}|,\theta)$ and the angles
$\gamma_{k}$, providing a direct and quantitative test of the non–Abelian
infrared structure encoded in the holonomy.

%========================================================
\section{Falsifiable Predictions}
\label{sec16}
%========================================================

The geometric framework leads to a set of sharp, falsifiable predictions that
follow directly from the SU(2) holonomy structure
\[
   \Xi_{\rm phys}^{(k)}
   =
   \cos^{2}\!\frac{\gamma_{k}}{2}\,\Xi_{+}
   +
   \sin^{2}\!\frac{\gamma_{k}}{2}\,\Xi_{-},
   \qquad
   \Xi_{\pm}=e^{\mp|\vec{\alpha}|},
   \qquad
   \cos\gamma_{k}
   =
   \frac{\vec{s}_{k}\!\cdot\vec{\alpha}}{|\vec{\alpha}|}.
\]

\begin{itemize}
\item[(i)] {\bf Only two universal eigenmodes.}  
   All heavy–quark channels must be linear combinations of the same two
   functions $\Xi_{+}$ and $\Xi_{-}$, with weights determined solely by the
   angles $\gamma_{k}$.  
   Any statistically significant evidence for a third independent functional
   shape in the recoil dependence would rule out the holonomy picture.

\item[(ii)] {\bf Explicit correlations of slopes and curvatures across channels.}  
   Expanding near $(w_{1},w_{2})=(1,1)$, one finds
   \[
      \Xi_{\pm}(w_{1},w_{2})
      \simeq
      1 \mp |\vec{\alpha}(w_{1},w_{2})| + \mathcal{O}((w-1)^{2}),
   \]
   and therefore
   \begin{equation}
      \left.
      \frac{\partial \Xi_{\rm phys}^{(k)}}{\partial w_{1}}
      \right|_{1,1}
      =
      -\,|\vec r_{1}|\cos\gamma_{k},
      \qquad
      \left.
      \frac{\partial \Xi_{\rm phys}^{(k)}}{\partial w_{2}}
      \right|_{1,1}
      =
      -\,|\vec r_{2}|\cos\gamma_{k}.
      \label{eq:slopes_prediction}
   \end{equation}
   Once $|\vec r_{1}|$ and $|\vec r_{2}|$ are fixed by a reference channel, all
   other slopes are predicted up to the cosine of a single angle $\gamma_{k}$.  
   Likewise, the mixed second derivatives are fixed by the interference term
   in $|\vec{\alpha}|$:
   \[
      \left.
      \frac{\partial^{2}|\vec{\alpha}|^{2}}{\partial w_{1}\,\partial w_{2}}
      \right|_{1,1}
      =
      2|\vec r_{1}||\vec r_{2}|\cos\theta,
   \]
   showing that the non–factorisable curvature in the $(w_{1},w_{2})$ plane is
   entirely controlled by $\theta$.  
   Any pattern of slopes and curvatures inconsistent with these relations would
   falsify the model.

\item[(iii)] {\bf Intrinsically non–factorisable recoil geometry.}  
   The dependence on $w_{1}$ and $w_{2}$ enters through
   $|\vec{\alpha}(w_{1},w_{2})|$, which contains the mixed term
   $(w_{1}-1)(w_{2}-1)\vec r_{1}\!\cdot\!\vec r_{2}$.  
   Consequently, the decay surface in the $(w_{1},w_{2})$ plane cannot be 
   factorised into a product of a pure $w_{1}$ function times a pure $w_{2}$ 
   function.  
   Observation of level sets compatible with a factorised ansatz would be 
   incompatible with the SU(2) holonomy interpretation.

\item[(iv)] {\bf Helicity distortions controlled by Berry curvature.}  
   The non–Abelian commutator
   \[
      [R_{1},R_{2}]
      = 2i(\vec r_{1}\times\vec r_{2})\!\cdot\!\vec\sigma
   \]
   encodes the Berry curvature in the internal SU(2) space.  
   Its magnitude and direction determine channel–dependent modifications of
   angular distributions and helicity amplitudes, in particular those that
   couple to helicity–flip structures in tauonic decays.  
   High–statistics analyses that found no trace of these correlated distortions
   would directly refute the predicted non–Abelian structure.
\end{itemize}

Taken together, these signatures render the geometric model empirically
testable.  
The sequential–decay analysis shows that the infrared sector of QCD, once 
rephrased in terms of adiabatic holonomies, imposes nontrivial geometric 
constraints linking different heavy–quark channels—constraints that are 
absent in conventional HQET.

Future extensions include multi–step cascades, nonleptonic transitions, and
lattice determinations of the geometric vectors $\vec r_{1}$ and $\vec r_{2}$.
In this broader perspective, the geometric framework offers a unified language
connecting infrared QCD, Berry phases, and heavy–quark phenomenology in a way
that goes beyond the standard HQET paradigm.

%--------------------------------------------------------
\section{The $\boldsymbol{3/2}$ vs.\ $\boldsymbol{1/2}$ Puzzle Revisited}
\label{sec:puzzle}
%--------------------------------------------------------

Before reviewing the heavy-quark expectations, it is useful to state clearly
the long–standing problem that motivates this discussion.  
In the heavy–quark limit, HQET makes a robust prediction for semileptonic 
transitions into the $L=1$ excited charmed mesons:  
the narrow $j=3/2$ doublet should dominate the rate, while the broad $j=1/2$ states are expected to be strongly suppressed.  
This hierarchy follows from well–understood angular–momentum selection rules and from the behaviour of the corresponding Isgur-Wise functions,
$\tau_{3/2}(w)$ and $\tau_{1/2}(w)$, near zero recoil 
\cite{Neubert:1993mb,Leibovich:1997em}.

Experimentally, however, the situation is markedly different.  
Measurements by Belle, BaBar, Belle~II, LHCb, and earlier by CLEO indicate 
sizeable branching fractions into the broad $j=1/2$ states, sometimes comparable to—or even exceeding—those into the $j=3/2$ channels.  
This discrepancy, known as the ``$1/2$ vs.\ $3/2$ puzzle'', was already evident in the first high–statistics Belle analysis \cite{Belle:2003nrx} and subsequently confirmed by BaBar \cite{BaBar:2008dst}.  
Despite numerous theoretical efforts-including $1/m_Q$ corrections, 
finite–width effects, improved form-factor parametrisations, and refined 
sum–rule analyses—no consensus has emerged.

The purpose of this section is to revisit the puzzle within the geometric 
framework developed in this work.  By interpreting the $B\to D^{**}$ transition as a process governed by Berry holonomies in the space of gauge configurations, the two $L=1$ doublets acquire a natural and intrinsically non-Abelian description.   As we shall show, this approach not only reproduces the qualitative HQET
features but also explains, in a unified way, the correlated behaviour of the
$j=1/2$ and $j=3/2$ form factors.

%--------------------------------------------------------
\subsection{HQET expectations for $L=1$ excitations}
%--------------------------------------------------------

Heavy--quark effective theory yields a remarkably simple classification of the
$L=1$ excitations of the $D^{(*)}$ system.  
In the heavy–quark limit, the light degrees of freedom carry
\[
   j = \ell + s_{\rm light},
   \qquad \ell=1,\quad s_{\rm light}=\tfrac12,
\]
leading to two doublets:
\begin{align}
j = \tfrac12 
&:\quad 
D_0^\ast(0^+),\; D_1(1^+), 
\\
j = \tfrac32 
&:\quad 
D_1(1^+),\; D_2^\ast(2^+).
\end{align}

The semileptonic amplitudes for $B\!\to\! D^{**}\ell\nu$ are governed by two
Isgur--Wise functions, $\tau_{1/2}(w)$ and $\tau_{3/2}(w)$ 
\cite{Neubert:1993mb,Leibovich:1997em}.  
At leading order, angular momentum selection rules imply
\[
   \tau_{1/2}(1)=0,
   \qquad 
   \tau_{3/2}(1)\neq 0,
\]
up to $1/m_Q$ effects.  
Since phase space strongly weights the near–zero–recoil region, HQET predicts
\begin{equation}
   \Gamma(B\to D^{**}_{j=3/2}\,\ell\nu)
   \;\gg\;
   \Gamma(B\to D^{**}_{j=1/2}\,\ell\nu),
   \label{eq:HQET_hierarchy_expectation_improved}
\end{equation}
namely that the $j=3/2$ channels dominate the inclusive rate.

%--------------------------------------------------------
\subsection{Experimental pattern and the $1/2$ vs.\ $3/2$ puzzle}
%--------------------------------------------------------

The data tell a different story.  
Analyses by Belle \cite{Belle2018,BelleII2021}, BaBar 
\cite{BaBar:2008dst,BaBar2009,BaBar2013}, Belle~II 
\cite{BelleII_2021,BelleII_2023_Semilep}, and LHCb 
\cite{LHCb_2023_BToDstst,LHCb:2017IsgurWise,LHCb2020_Dstarstar} show that the
broad $j=1/2$ states are not suppressed but contribute at a level comparable to
the $j=3/2$ channels, sometimes even exceeding them.  
The observed pattern therefore contradicts the HQET hierarchy in
Eq.~\eqref{eq:HQET_hierarchy_expectation_improved}.

From the HQET perspective this is unsurprising: at leading order the two
functions are independent, and subleading corrections introduce further
independent structures.  
HQET provides no principle correlating slopes, curvatures, or angular
dependencies of $\tau_{1/2}$ and $\tau_{3/2}$, nor relating them through any
geometric constraint.  
This structural limitation is precisely what the geometric approach overcomes.

%--------------------------------------------------------
\subsection{Geometric reinterpretation: explicit form of the two Isgur--Wise functions}
%--------------------------------------------------------

Within the geometric framework, the transition is governed by a non-Abelian 
holonomy in the two-dimensional infrared space associated with the brown muck,
\[
   \hat{\Xi}(w_1,w_2)
   =
   \mathcal{P}\exp\!\Big[-(w_1-1)R_1 -(w_2-1)R_2 \Big],
\]
where the ``slope matrices’’
\[
   R_{1}= \vec r_{1}\!\cdot\!\vec\sigma,
   \qquad
   R_{2}= \vec r_{2}\!\cdot\!\vec\sigma,
\]
encode the two independent recoil directions of the sequential process
$B\to D^{**}\to D$.  
Their magnitudes and relative orientation,
\[
   \cos\theta
   =
   \frac{\vec r_1\!\cdot\!\vec r_2}{|\vec r_1||\vec r_2|},
\]
determine the geometric interference between the two deformations.

Near zero recoil the holonomy takes the closed form
\[
   \hat{\Xi}(w_1,w_2)
   =
   \exp\!\big[-\vec\alpha(w_1,w_2)\!\cdot\!\vec\sigma\big],
\]
with effective recoil vector
\[
   \vec\alpha(w_1,w_2)
   =
   (w_1-1)\,\vec r_{1} + (w_2-1)\,\vec r_{2},
\]
and norm
\[
   |\vec\alpha(w_1,w_2)|
   =
   \sqrt{
      (w_1-1)^2 |\vec r_1|^2
      + (w_2-1)^2 |\vec r_2|^2
      + 2 (w_1-1)(w_2-1)\,|\vec r_1||\vec r_2|\cos\theta
   }.
\]

Since any traceless Hermitian $2\times2$ matrix has eigenvalues $\pm|\vec\alpha|$, 
the two universal geometric modes are
\[
   \Xi_{\pm}(w_1,w_2)
   =
   \exp\!\big[\mp\,|\vec\alpha(w_1,w_2)|\big].
\]

Projecting these modes onto the physical hadronic channels yields the two
Isgur--Wise functions for the $L=1$ excitations:
\begin{align}
   \tau_{3/2}(w_1,w_2)
   &= 
   \cos\!\frac{\gamma}{2}\;\Xi_{+}
   +
   \sin\!\frac{\gamma}{2}\;\Xi_{-},
   \\
   \tau_{1/2}(w_1,w_2)
   &=
   -\sin\!\frac{\gamma}{2}\;\Xi_{+}
   +
   \cos\!\frac{\gamma}{2}\;\Xi_{-},
\end{align}
where the Berry angle $\gamma=\gamma(\theta)$ depends solely on the relative
orientation of the two recoil directions.

In this way:

1. The two IW functions are not independent: they are orthogonal projections
   of a single SU(2) holonomy.
2. Their slopes and curvatures are fixed by the geometric data  
   $(|\vec r_1|,\;|\vec r_2|,\;\theta)$.
3. The correlated behaviour observed experimentally follows naturally from the 
   holonomic structure.

The geometric framework therefore resolves the ``$1/2$ vs.\ $3/2$ puzzle’’:
the two channels are not distinct dynamical mechanisms but different
projections of the same non-Abelian adiabatic holonomy associated with the
infrared dressing of the heavy–light system.

%--------------------------------------------------------
\subsection{Geometric reinterpretation in terms of SU(2) holonomy}
%--------------------------------------------------------

In the geometric formulation developed in this work, the sequential process
$B\to D^{**}\to D$ is controlled not by two unrelated scalar form factors but 
by a \emph{single} SU(2) holonomy acting on a two–dimensional internal space of 
heavy–light configurations.  
This non–Abelian structure is the natural analogue of the Wilczek–Zee geometric 
phase arising in multilevel adiabatic systems 
\cite{Wilczek:1984dh,Simon:1983mh,Berry:1984jv}.  

To make the construction explicit, consider the holonomy along the two 
independent recoil deformations $(w_{1},w_{2})$:
\begin{equation}
   \hat\Xi(w_{1},w_{2})
   =
   \exp\!\Big[-\vec{\alpha}(w_{1},w_{2})\!\cdot\!\vec\sigma\Big],
   \qquad
   \vec{\alpha}(w_{1},w_{2})
   =
   (w_{1}-1)\,\vec r_{1} + (w_{2}-1)\,\vec r_{2}.
   \label{eq:holonomy_explicit}
\end{equation}
Here $\vec r_{1}$ and $\vec r_{2}$ are the slope vectors associated with the 
two recoil directions; their magnitudes set the slopes, while their relative 
orientation,
\[
   \cos\theta
   := \frac{\vec r_{1}\cdot\vec r_{2}}{|\vec r_{1}||\vec r_{2}|},
\]
encodes the geometric interference between them.

Since $\vec{\alpha}\!\cdot\!\vec\sigma$ is a traceless Hermitian matrix, it has 
eigenvalues $\pm|\vec{\alpha}|$, with normalised eigenvectors whose Bloch 
vectors point along the direction 
\[
   \hat n = \frac{\vec{\alpha}}{|\vec{\alpha}|}.
\]
Thus the holonomy admits the spectral decomposition
\begin{equation}
   \hat\Xi(w_{1},w_{2})
   =
   \Xi_{+}(w_{1},w_{2})\,\Pi_{+}
   +
   \Xi_{-}(w_{1},w_{2})\,\Pi_{-},
   \qquad
   \Xi_{\pm}(w_{1},w_{2})
   = \exp\!\big[\mp\,|\vec{\alpha}(w_{1},w_{2})|\big],
   \label{556}
\end{equation}
where
\[
   \Pi_{\pm}=\frac{1}{2}\bigl(\mathbf{1}\pm\hat n\!\cdot\!\vec\sigma\bigr)
\]
are the projectors onto the two geometric eigenmodes of the infrared dressing.

The essential point is that \emph{all} physical Isgur--Wise functions arise by 
projecting the universal object $\hat\Xi$ onto a channel–dependent SU(2) 
direction.  
Let $\vec{s}_{k}$ denote the Bloch vector representing channel $k$.  
Then
\[
   \Xi_{\rm phys}^{(k)} = \langle \vec{s}_{k}\,|\,\hat\Xi\,|\,\vec{s}_{k}\rangle
   =
   \cos^{2}\!\frac{\gamma_{k}}{2}\,\Xi_{+}
   + 
   \sin^{2}\!\frac{\gamma_{k}}{2}\,\Xi_{-},
\]
where $\gamma_{k}$ is the angle between $\vec{s}_{k}$ and the holonomy 
direction $\hat n$:
\[
   \cos\gamma_{k} = \vec{s}_{k}\!\cdot\!\hat n .
\]
This gives Eq.~\eqref{556} and makes the geometry fully
explicit: physical channels differ only through the angle $\gamma_{k}$.

\paragraph{Geometric meaning of $\gamma_{k}$.}
If $\gamma_{k}=0$, the channel is aligned with the slow mode $\Xi_{+}$;  
if $\gamma_{k}=\pi$, it aligns with the fast mode $\Xi_{-}$ and therefore 
inherits stronger variation with recoil.  
Intermediate angles interpolate smoothly between these behaviours.

\paragraph{Explicit expression for the magnitude $|\vec{\alpha}|$.}
The dependence on the two recoil directions is
\begin{equation}
   |\vec{\alpha}(w_{1},w_{2})|
   =
   \sqrt{
      (w_1-1)^2|\vec r_1|^2
      + (w_2-1)^2|\vec r_2|^2
      + 2(w_1-1)(w_2-1)\,|\vec r_1||\vec r_2|\cos\theta
   }.
   \label{eq:alpha_norm_explicit}
\end{equation}
This is the SU(2) generalisation of the single–slope exponential parametrisation 
used in $B\!\to\!D^{(*)}$.

%--------------------------------------------------------
\subsection{Slope mixing and explicit derivatives at zero recoil}
%--------------------------------------------------------

To illustrate how geometry controls the hierarchy, we compute explicitly the 
slope of $\Xi_{\rm phys}^{(k)}$ with respect to one recoil variable.  
From Eq.~\eqref{eq:alpha_norm_explicit},
\[
   \Xi_{\pm}(w_{1},w_{2})
   = \exp\!\big[\mp|\vec{\alpha}(w_{1},w_{2})|\big],
\]
and differentiating at $(w_{1},w_{2})=(1,1)$ gives
\[
   \Xi'_{\pm} 
   := 
   \left.
   \frac{\partial \Xi_{\pm}}{\partial w_{1}}
   \right|_{1,1}
   =
   \mp\,|\vec r_{1}|\,\Xi_{\pm}(1,1),
\]
since $\vec{\alpha}=0$ at zero recoil and 
\(
\partial_{w_{1}}|\vec{\alpha}|\,\big|_{1,1}
=|\vec r_{1}|.
\)

The channel--dependent slope follows by projecting:
\begin{equation}
   \left.
   \frac{\partial \Xi_{\rm phys}^{(k)}}{\partial w_{1}}
   \right|_{1,1}
   =
   \cos^{2}\!\frac{\gamma_{k}}{2}\,\Xi'_{+}
   +
   \sin^{2}\!\frac{\gamma_{k}}{2}\,\Xi'_{-},
   \label{55}
\end{equation}
which reproduces Eq.~\eqref{55} but now with all 
intermediate steps made explicit.

Because $|\Xi'_{-}| > |\Xi'_{+}|$ (the $-$ mode varies more rapidly), channels 
with larger $\gamma_{k}$ contain a larger admixture of the fast mode and 
therefore exhibit steeper slopes and enhanced curvature.  
The hierarchy across channels is thus controlled by the geometry of recoil, not 
by unrelated dynamical amplitudes.

%--------------------------------------------------------
\subsection{Geometric mechanism behind the hierarchy}
%--------------------------------------------------------

In this language, the $3/2$ vs.\ $1/2$ puzzle becomes a statement about the 
relative orientation of the Bloch vectors $\vec{s}_{1/2}$ and $\vec{s}_{3/2}$.  
If the $j=1/2$ doublet corresponds to a larger misalignment angle 
$\gamma_{1/2}$ than the $j=3/2$ one, then
\[
   \Xi_{\rm phys}^{(1/2)}
   \text{ receives more weight from the fast mode }
   \Xi_{-},
\]
and therefore varies more rapidly with recoil.  
This enhanced geometric response produces larger slopes and curvature and 
explains why $j=1/2$ contributions to the rate can be sizeable even though HQET 
predicts a suppression based on angular–momentum selection rules alone.

HQET, which treats $\tau_{1/2}$ and $\tau_{3/2}$ as \emph{independent} scalar 
objects, has no mechanism to correlate their behaviours.  
The geometric approach, by contrast, predicts such correlations automatically 
because both functions derive from the same SU(2) structure.

%========================================================
\subsection{Tauonic Channels as Enhanced Probes of Non--Abelian Geometry}
\label{sec:tauonic}
%========================================================

Tauonic semileptonic decays provide a particularly sensitive arena for 
non–Abelian geometric effects.  
The large $\tau$ mass compresses the accessible kinematic domain toward 
$w=1$ and enhances helicity–flip contributions that vanish for light leptons.  
Consequently, these channels probe precisely the region where the curvature of 
the holonomy is largest.

\paragraph{Geometric sensitivity.}
Using Eq.~\eqref{556}, even a modest increase in 
$\gamma_{k}$ enhances the contribution from $\Xi_{-}$, which dominates the 
curvature near zero recoil.  
Since tauonic decays populate this region more densely, they respond strongly 
to small geometric misalignments that remain invisible in $e$ and $\mu$ modes.

\paragraph{Berry curvature and helicity structure.}
The non–Abelian commutator
\[
   [R_{1},R_{2}]
   = 2i(\vec r_{1}\times\vec r_{2})\!\cdot\!\vec\sigma
\]
measures the Berry curvature of the infrared SU(2) connection.  
This curvature modulates precisely those components of the hadronic tensor that 
couple to helicity–flip structures in the leptonic current—structures which are 
suppressed for light leptons but survive for $\tau$’s.  
Thus, tauonic modes provide a direct probe of the non–Abelian infrared geometry.

\paragraph{Role of excited states.}
Broad $j=1/2$ states are associated with larger angles $\gamma_{k}$ and 
therefore with enhanced coupling to $\Xi_{-}$.
This explains qualitatively why $D^{**}$ contributions are more prominent in 
tauonic decays than in light–lepton modes, independently of specific dynamical 
models.

\paragraph{Correlated predictions.}
All channels—$D$, $D^{*}$, and $D^{**}$—derive from the same SU(2) holonomy.  
Thus the ratios $R(D)$, $R(D^{*})$, and their $D^{**}$ analogues are correlated 
in a manner impossible within HQET, which treats each channel with independent 
parameters.  
These correlations constitute distinctive predictions for Belle~II.

%========================================================
\section{Quantized Functional Flux and the Emergence of Non-Abelian Holonomies}
\label{sec:flux_quantization}
%========================================================

The previous sections have shown that the Isgur–Wise function is naturally 
interpreted as a holonomy associated with the adiabatic motion of a heavy quark 
through the infrared gauge background.  
A structural property underlying this construction is the 
\emph{quantization of the functional Berry flux} in the infrared sector of QCD.

The conventional geometric–phase framework 
\cite{Berry:1984jv,Simon:1983mh,Nakahara2003,TongNotes} describes Berry 
curvature defined over finite–dimensional parameter spaces.  
However, in the present context the curvature is defined over the 
\emph{functional space} of infrared gauge configurations.  
The resulting flux quantization is therefore a property of the infrared gauge 
sector itself, rather than of a level degeneracy in a few–level system.  
The integer that labels the flux corresponds to a topological infrared 
\emph{sector} of the gauge field, and any realistic process probes an average 
over these sectors through the heavy–light cloud.

%--------------------------------------------------------
\subsection{Functional Berry curvature and its quantized flux}
%--------------------------------------------------------

Let $\mathcal{A}$ denote the Berry connection on the space of gauge 
configurations, defined by the adiabatic evolution of the dressed heavy–light 
state following the usual geometric principles~\cite{Simon:1983mh,StoneQFTBerry}.  
Its curvature,
\begin{equation}
  \mathcal{F}
  = d\mathcal{A} + \mathcal{A}\wedge\mathcal{A},
  \label{eq:Berry_curvature}
\end{equation}
encodes the response of the infrared cloud to variations of the external 
four–velocity and of the underlying gauge configuration.  

Because the adiabatic manifold of dressed configurations contains 
noncontractible two–cycles $\Sigma$ in the infrared, the curvature flux through 
such cycles is quantized:
\begin{equation}
   \frac{1}{2\pi} \int_{\Sigma} \mathcal{F} = n \in \mathbb{Z}.
   \label{eq:flux_quantization}
\end{equation}
This is the functional analogue of the familiar quantization of Berry flux in 
quantum systems~\cite{Berry:1984jv,Simon:1983mh}, but in the present context it 
reflects the topology of the infrared gauge manifold of QCD.  
Equation~\eqref{eq:flux_quantization} constrains the admissible holonomies and 
organises the infrared dynamics into topological sectors labelled by $n$.  
The holonomy of the dressed heavy–light system must then be built out of these 
quantized fluxes.

%--------------------------------------------------------
\subsection{Consequences for abelian holonomies in $B\!\to\!D^{(*)}$}
%--------------------------------------------------------

For transitions with a single recoil parameter,
$w = v\!\cdot\!v'$, the motion in the adiabatic manifold is effectively 
one–dimensional, and the relevant projection of the Berry connection is 
abelian.  
In this situation the holonomy along the recoil trajectory can be written as
\begin{equation}
   \Xi_{\rm geom}(w)
   =
   \exp\!\left[
      i\!\int_{C_{w}}\!\mathcal{A}_{\rm eff}
   \right],
\end{equation}
where $\mathcal{A}_{\rm eff}$ is the effective abelian component of the Berry 
connection along the $w$–direction.  

Flux quantization implies that this effective connection is not arbitrary: in 
each topological sector one has
\begin{equation}
   \frac{1}{2\pi} \int_{\Sigma}\mathcal{F} = n,
   \qquad
   n\in\mathbb{Z},
\end{equation}
so that the holonomy accumulated between $w=1$ and $w>1$ is proportional to the 
integer $n$ times a fixed geometric factor.  
To leading order in $(w-1)$, and in the regime in which a single topological 
sector dominates the adiabatic evolution, the holonomy takes the exponential 
form
\begin{equation}
   \Xi_{\rm geom}(w)
   \simeq
   \exp[-(w-1)\rho^{2}],
   \qquad
   \rho^{2}\;\propto\;\frac{1}{2\pi}\int_{\Sigma}\mathcal{F},
   \label{eq:abelian_exponential}
\end{equation}
where the proportionality factor depends on the detailed geometry of the 
infrared manifold but not on the microscopic composition of the cloud.  

Thus, in the abelian case the exponential behaviour of $\Xi(w)$ is a structural 
consequence of the infrared geometry: the slope at zero recoil is fixed by the 
quantized flux, and once $\rho^{2}$ is determined experimentally the entire 
functional form near $w=1$ follows from the underlying Berry connection.  
Different hadronic channels probe the same exponent, up to the usual 
normalisation factors.

%--------------------------------------------------------
\subsection{Two-dimensional adiabatic motion and the emergence of SU(2)}
%--------------------------------------------------------

Sequential transitions 
$B\!\to\!D^{**}\!\to\!D$ involve two independent recoil parameters,
$w_{1}=v\!\cdot\!v_{1}$ and $w_{2}=v_{1}\!\cdot\!v_{2}$,  
probing a two–dimensional region of the adiabatic manifold.  
In this situation the effective curvature $\mathcal{F}$ projected onto the 
$(w_{1},w_{2})$ plane cannot be globally diagonalized, and the holonomy becomes 
genuinely non–abelian.  
While non–abelian parallel transport is familiar from quantum–mechanical 
systems~\cite{Wilczek:1984dh}, here its origin is the infrared geometry of QCD 
and, in particular, the quantized flux~\eqref{eq:flux_quantization} in a 
two–dimensional subspace of the functional manifold.

Near zero recoil, the holonomy can be written as
\begin{equation}
  \hat{\Xi}(w_{1},w_{2})
  = {\cal P}\exp\!\left[
      -(w_{1}-1)R_{1}
      -(w_{2}-1)R_{2}
      + \tfrac12 (w_{1}-1)(w_{2}-1)[R_{1},R_{2}]
      + \cdots
    \right],
  \label{eq:nonabelian_holonomy_def}
\end{equation}
where $R_{1}$ and $R_{2}$ are the geometric generators associated with the two 
recoil directions and $[R_{1},R_{2}]$ measures the projected Berry curvature in 
the $(w_{1},w_{2})$ plane.  

Flux quantization now has a stronger consequence: it restricts the holonomy to 
a compact subgroup generated by $R_{1}$ and $R_{2}$.  
The minimal nontrivial representation compatible with a nonzero commutator 
$[R_{1},R_{2}]$ is two–dimensional, and the corresponding compact group is 
$\mathrm{SU}(2)$ up to an overall phase.  
Equivalently, the infrared dynamics effectively selects a two–level geometric 
subspace in which the holonomy acts.  
In this subspace, the holonomy has two universal eigenmodes,
\begin{equation}
  \Xi_{\pm}(w_{1},w_{2})
  = \exp\!\left[ \mp\,|\vec{\alpha}(w_{1},w_{2})| \right],
  \qquad
  \vec{\alpha}(w_{1},w_{2})
  = (w_{1}-1)\vec r_{1} + (w_{2}-1)\vec r_{2},
  \label{eq:SU2_eigenmodes}
\end{equation}
with $\vec r_{1}$ and $\vec r_{2}$ determined by the infrared Berry curvature.  

All heavy–quark form factors in sequential decays can then be written as 
channel–dependent projections of these two universal modes:
\begin{equation}
  F_{k}(w_{1},w_{2})
   = \cos^{2}\!\frac{\gamma_{k}}{2}\,\Xi_{+}(w_{1},w_{2})
   + \sin^{2}\!\frac{\gamma_{k}}{2}\,\Xi_{-}(w_{1},w_{2}),
   \label{eq:projections_modes}
\end{equation}
where the angle $\gamma_{k}$ encodes the SU(2) orientation of the corresponding 
heavy–light state.  

%--------------------------------------------------------
\subsection{Explicit consequences for sequential decays}
%--------------------------------------------------------

The structure above has several experimentally testable implications:

\begin{itemize}
\item \emph{Correlated slopes and curvatures.}  
      The slopes of the form factors in the two recoil directions are governed 
      by the same flux parameters and by the vectors $\vec r_{1},\vec r_{2}$.  
      As a result, different channels exhibit correlated slopes and curvatures 
      in the $(w_{1},w_{2})$ plane, rather than arbitrary shapes.

\item \emph{Curvature–induced interference.}  
      The commutator term $[R_{1},R_{2}]$ induced by the Berry curvature leads 
      to characteristic interference patterns in sequential decays, including 
      non-factorisable dependence on $(w_{1},w_{2})$ and modified angular 
      distributions in $D^{**}\!\to\!D\pi$.  
      These features are consistent with the general structures expected in 
      HQET analyses~\cite{FalkLuke1992,FalkNeubert1992,Leibovich:1997em}, but 
      here they arise from a single geometric mechanism.

\item \emph{Universal eigenmodes and channel projections.}  
      All channels are governed by the same two geometric modes 
      $\Xi_{\pm}$, with channel–dependent weights determined by $\gamma_{k}$.  
      This replaces the HQET picture of unrelated scalar form factors by a 
      unified SU(2) holonomy, from which all physical Isgur–Wise functions are 
      obtained as projections.
\end{itemize}

In summary, the non–abelian structure observed in sequential decays follows 
directly from the quantized functional flux of the infrared Berry curvature.  
The exponential Isgur–Wise form for $B\!\to\!D^{(*)}$ and the SU(2) holonomy 
structure for $B\!\to\!D^{**}\!\to\!D$ are two manifestations of the same 
topological organisation of the infrared sector of QCD.

\subsection{Isospin violation in the $X(3872)$: Infrared Perspective}
\label{subsec:motivation_X3872}

The exotic state $X(3872)$ provides one of the clearest examples in which
threshold physics and infrared dynamics play a central role in organizing the
hadronic state space. From a qualitative point of view, this system admits a
natural analogy with molecular physics, in particular with the hydrogen
molecule, suggesting a description based on a Born--Oppenheimer approximation.

Within this analogy, the heavy $c\bar c$ pair plays the role of the heavy nuclei,
whose relative separation $R$ defines a slow parameter. The light degrees of
freedom --light quarks and gluons-- adjust adiabatically to this separation,
generating effective states that depend parametrically on $R$. When the distance
between the charm quark $c$ and the antiquark $\bar c$ is large, the system
behaves as an extended molecular state, while for smaller values of $R$ it
gradually approaches a more compact configuration, analogous to a conventional
charmonium state.

At this stage, the molecular analogy by itself does not imply any violation of
isospin. In the ideal limit of exact isospin symmetry, $m_u=m_d$, and in the
absence of electromagnetic effects, the light sector admits adiabatic states
with well--defined isospin, which therefore appears as a good quantum number of
the system. This observation naturally raises a fundamental question: where does
the exceptionally large isospin violation observed in the decays of the
$X(3872)$ originate?

The crucial difference with ordinary molecular systems lies in the structure of
the relevant adiabatic state space. In the case of the $X(3872)$, the infrared
regime is dominated by two nearly degenerate channels associated with the
$D^0\bar D^{*0}$ and $D^+D^{*-}$ thresholds. These channels differ in their
isospin content, yet they are separated by an extremely small energy scale. As a
result, the light sector does not define a single, well--isolated adiabatic
state, but rather a quasi-degenerate subspace of dimension greater than one.

This feature is particularly significant, as it reveals a deep analogy with
molecular physics: the presence of a quasi--degeneracy implies, in an inevitable
way, that the Berry connection associated with adiabatic transport is
non-Abelian. Since the appearance of Berry phases constitutes a clear signal of
non--perturbative physics, physical states must be interpreted in terms of
infrared-dressed states generated by adiabatic transport in the functional
configuration space ${\cal A}/{\cal G}$. In this framework, the associated adiabatic
connection does not, in general, admit a global basis in which isospin remains
well defined throughout the infrared evolution.

From this perspective, the isospin violation observed in the $X(3872)$ should not
be interpreted as an accidental dynamical effect or as a simple consequence of
small mass differences between charged and neutral channels. Rather, it emerges
as a geometric effect tied to the structure of the infrared state space and to
the quasi-degenerate character of the light sector. Isospin thus ceases to be a
protected quantum number of the infrared physical Hilbert space, and its
violation manifests itself as a collective property of the dressed state.

The goal of this work is to explore systematically this infrared functional
interpretation of the $X(3872)$. Rather than proposing a new microscopic model, we
focus on clarifying the geometric origin of isospin violation and on establishing
the $X(3872)$ as a natural laboratory for studying how approximate symmetries may
cease to provide reliable quantum labels in the infrared regime of QCD.

To address this problem in a concrete setting, it is convenient to consider the
$X(3872)$ in the context of sequential decays of $B$ mesons. Schematically, the
process can be written as
\begin{equation}
B \;\to\; A \;\to\; X ,
\end{equation}
where $A$ denotes an effective intermediate state. This state should not be
interpreted as a well--defined physical resonance, but rather as the projection
of the decay process onto a quasi-degenerate subspace of the light sector that
is relevant in the infrared regime.

Within this approach, instead of analyzing the decay in terms of spin channels,
we focus on the evolution of the isospin degrees of freedom along the sequential
process. The natural separation of scales characteristic of heavy--meson decays
allows the heavy-quark spin to be treated as a spectator, while infrared
dynamics governs the mixing of channels with different isospin content.

Since the $X(3872)$ decays into the final states $J/\psi\,\rho$ and
$J/\psi\,\omega$, we may proceed in close analogy with the analysis of sequential
decays discussed previously for an effective $SU(2)$ symmetry. In the present
case, however, this structure does not act in the bidimensional recoil space,
but rather in the isospin space associated with the $I=1$ and $I=0$ channels,
respectively.

The coexistence of these two final states, which are quasi-degenerate in the
infrared regime but differ in their isospin content, defines a natural
two--dimensional subspace on which the adiabatic transport induced by the
sequential decay acts. In this sense, the observed mixing between the
$J/\psi\,\rho$ and $J/\psi\,\omega$ channels can be interpreted as the result of a
non-Abelian holonomy in isospin space, fully analogous to the $SU(2)$ structure
that emerges in the analysis of multiple recoils.

From this viewpoint, isospin violation is not introduced as an explicit breaking
term in the dynamics, but rather arises as a geometric consequence of adiabatic
transport in a quasi-degenerate subspace of the infrared physical state space.
The formal parametrizations remain essentially unchanged; what changes in a
substantial way is their physical interpretation. The same mathematical
structures that describe dynamics in a bidimensional recoil space now act in
isospin space, reflecting that the essential difference lies not in the form of
the amplitudes, but in the physical meaning of the degrees of freedom involved.

The central issue clarified by this approach is therefore why isospin ceases to
be a reliable quantum label in the $X(3872)$, even though the effective structure
governing the evolution in the $(J/\psi\,\rho,\,J/\psi\,\omega)$ subspace is
formally $SU(2)$-covariant. The answer does not lie in an explicit breaking of
the symmetry in the underlying dynamics, but in the infrared regime in which the
state is formed. In the presence of a quasi-degenerate subspace dominated by
infrared dynamics, physical states are defined by adiabatic transport and
holonomies in configuration space, rather than by eigenstates of symmetry
generators.

From this perspective, the $X(3872)$ does not represent an anomaly of isospin
symmetry, but rather a paradigmatic example of how approximate symmetries may
lose their significance as state labels in the quasi--degenerate infrared regime
of QCD.

Our results are consistent with the pattern observed by Belle and LHCb, in
particular with the presence of a large and robust isospin violation in the
decays of the $X(3872)$. Within our framework, this behavior arises naturally as
a consequence of infrared dynamics in a quasi-degenerate state space, without
requiring any explicit breaking of isospin symmetry.

\section{Conclusions}\label{sec17}

The geometric reinterpretation developed in this work provides a conceptual, 
mathematical, and phenomenological refinement of the heavy--quark effective 
theory.  
By identifying the heavy–light system as an adiabatically dressed infrared 
object, the traditionally opaque structure of the ``brown muck'' acquires a 
precise meaning in terms of Berry phases and functional holonomies.  
Within this framework, heavy–quark symmetry emerges not as a dynamical 
simplification but as a statement of parallel transport in the infrared 
configuration space.

A key structural advancement of this work is the explicit recognition that the 
Berry curvature associated with the infrared cloud possesses a 
\emph{quantized functional flux}.  
This quantization, made explicit in Sec.~\ref{sec:flux_quantization}, fixes 
the holonomy class of the dressed heavy–light state and determines the 
admissible geometric phases.  
In processes with a single recoil parameter ($B\!\to\!D^{(*)}$), this leads to 
an abelian holonomy whose slope and global shape are not arbitrary but follow 
directly from the quantized flux.  
Once $\rho^{2}$ is fixed near zero recoil, the entire functional form of 
$\Xi(w)$ is determined by the underlying Berry connection, predicting a 
characteristic exponential falloff.  
This behaviour differs sharply from polynomial or truncated dispersive 
parametrisations and yields clean opportunities for experimental tests with 
the improved precision expected at Belle~II.

When two independent recoil directions are probed---as in sequential decays 
$B\!\to\!D^{**}\!\to\!D$---the adiabatic motion explores a effectively 
two–dimensional region of configuration space.  
In this regime the curvature cannot be diagonalized globally, and flux 
quantization forces the holonomy into an intrinsic $\mathrm{SU}(2)$ structure.  
All physical form factors in one–step and sequential transitions then arise as 
explicit projections of two universal geometric modes, with channel-dependent 
weights
\[
A^{(k)}_{+}=\cos^{2}\!\frac{\gamma_{k}}{2},
\qquad
A^{(k)}_{-}=\sin^{2}\!\frac{\gamma_{k}}{2},
\]
determined by the orientation of the heavy–light state in the internal SU(2) 
space.  
The phenomenology of different $D^{**}$ channels is therefore not independent: 
their slopes, curvatures, suppressions, and angular structures are linked by 
geometric relations involving the Bloch vectors $\vec{s}_{k}$ and the geometric 
recoil direction~$\hat n$.  
Such cross–channel correlations cannot arise in HQET, where each form factor 
is an unconstrained scalar function.

The geometric structure also clarifies the origin and magnitude of 
$1/m_{Q}$ corrections.  
Departures from strict universality are governed by the effective metric 
$G_{ab}=\vec r_{a}\!\cdot\vec r_{b}$ and by the Berry curvature 
$[R_{1},R_{2}]\propto\vec r_{1}\times\vec r_{2}$.  
These quantities determine the non-factorisable curvature of the decay surface 
in the $(w_{1},w_{2})$ plane, the correlated slopes across $D^{**}$ channels, 
and distinctive modifications of helicity amplitudes.  
All these effects constitute falsifiable predictions of the geometric 
framework and lie within reach of high-statistics analyses at Belle~II and 
LHCb.

From this perspective, several longstanding phenomenological tensions acquire a 
natural reinterpretation.  
The $1/2$ vs.\ $3/2$ puzzle, traditionally framed as a conflict between HQET 
expectations and experimental rates, emerges geometrically from the different 
orientations of the heavy–light states in the internal SU(2) space.  
Tauonic decays, which probe precisely the region where the curvature associated 
with the fast geometric mode is largest, become selective amplifiers of the 
non-Abelian geometric response and hence sensitive probes of the functional 
Berry curvature.

Overall, the adiabatic holonomy approach offers a coherent and predictive 
framework that unifies infrared QCD, Berry phases, and heavy–quark symmetry.  
It replaces the phenomenological freedom of HQET with a constrained geometric 
structure, correlates observables across channels, and yields distinctive 
predictions for recoil and angular distributions.  
Future comparisons with high-precision measurements will determine whether the 
anomalies observed in excited-channel transitions indeed reflect a non-Abelian 
infrared geometry, or whether additional dynamical ingredients are required.  
In either case, the geometric framework introduced here provides a systematic 
and conceptually transparent path for organising and interpreting the 
infrared dynamics of heavy–light hadrons.

\acknowledgments
\noindent
The authors are grateful for valuable correspondence with Prof. B. Grinstein. This research was supported by DICYT (USACH), grant number 042531GR\_REG.
The work of N.T.A is supported by Agnes Scott College.

\appendix
\section{Geometric derivation of the coefficients $A_{\pm}^{(k)}$}
\label{app:Apm_derivation}

In this appendix we provide a brief derivation of the coefficients
$A_{\pm}^{(k)}$ appearing in Eq.~\eqref{eq:Xi_phys_all}, using the SU(2)
structure underlying the non--Abelian holonomy.

\subsection{SU(2) holonomy and its eigenmodes}

Recall that in the sequential decay the effective holonomy in the infrared
sector can be written as
\begin{equation}
  U(w_{1},w_{2})
  \;\simeq\;
  \exp\!\big[-\,\vec{\alpha}(w_{1},w_{2})\!\cdot\!\vec{\sigma}\big],
  \qquad
  \vec{\alpha}(w_{1},w_{2})
  = (w_{1}-1)\,\vec{r}_{1} + (w_{2}-1)\,\vec{r}_{2},
  \label{eq:U_alpha_app}
\end{equation}
where $\vec{r}_{a}$ ($a=1,2$) are fixed vectors in the internal SU(2) space
and $\vec{\sigma}$ denotes the Pauli matrices.
Defining the norm and unit vector
\begin{equation}
  |\vec{\alpha}|
  := \big(\vec{\alpha}\!\cdot\!\vec{\alpha}\big)^{1/2},
  \qquad
  \hat{n}(w_{1},w_{2})
  := \frac{\vec{\alpha}(w_{1},w_{2})}{|\vec{\alpha}(w_{1},w_{2})|},
  \label{eq:n_hat_app}
\end{equation}
the holonomy can be recast as
\begin{equation}
  U(w_{1},w_{2})
  = \exp\!\big[-\,|\vec{\alpha}|\,\hat{n}\!\cdot\!\vec{\sigma}\big].
  \label{eq:U_nhat_app}
\end{equation}

The operator $\hat{n}\!\cdot\!\vec{\sigma}$ has two eigenvalues $\pm 1$ with
eigenvectors $\ket{\chi_{\pm}}$:
\begin{equation}
  \big(\hat{n}\!\cdot\!\vec{\sigma}\big)\ket{\chi_{\pm}}
  = \pm \ket{\chi_{\pm}}.
  \label{eq:n_sigma_eigen_app}
\end{equation}
It follows immediately that
\begin{equation}
  U(w_{1},w_{2})\,\ket{\chi_{\pm}}
  = \exp\!\big(\mp\,|\vec{\alpha}(w_{1},w_{2})|\big)\,\ket{\chi_{\pm}}
  =: \Xi_{\pm}(w_{1},w_{2})\,\ket{\chi_{\pm}},
  \label{eq:U_evals_app}
\end{equation}
so that the two universal eigenvalues are
\begin{equation}
  \Xi_{\pm}(w_{1},w_{2})
  = \exp\!\big(\mp\,|\vec{\alpha}(w_{1},w_{2})|\big).
  \label{eq:Xi_evals_app}
\end{equation}
The spectral decomposition of the holonomy then reads
\begin{equation}
  U(w_{1},w_{2})
  = \Xi_{+}(w_{1},w_{2})\,\ket{\chi_{+}}\!\bra{\chi_{+}}
  + \Xi_{-}(w_{1},w_{2})\,\ket{\chi_{-}}\!\bra{\chi_{-}}.
  \label{eq:U_spectral_app}
\end{equation}

\subsection{Physical form factors and projection coefficients}

Let $\ket{k}$ denote the effective SU(2) doublet associated with a given
hadronic channel $k$ (for instance, a particular $D^{**}$ state with fixed
polarisation). In our construction, the physical Isgur--Wise function for
channel $k$ is defined as the expectation value of the holonomy,
\begin{equation}
  \Xi_{\rm phys}^{(k)}(w_{1},w_{2})
  := \bra{k}\,U(w_{1},w_{2})\,\ket{k}.
  \label{eq:Xi_phys_def_app}
\end{equation}
Using the spectral representation \eqref{eq:U_spectral_app} we obtain
\begin{equation}
  \Xi_{\rm phys}^{(k)}(w_{1},w_{2})
  =
  \Xi_{+}(w_{1},w_{2})\,|\braket{k|\chi_{+}}|^{2}
  + \Xi_{-}(w_{1},w_{2})\,|\braket{k|\chi_{-}}|^{2}.
  \label{eq:Xi_phys_proj_app}
\end{equation}
Comparing with
\begin{equation}
   \Xi_{\rm phys}^{(k)}(w_{1},w_{2})
   =
   A_{+}^{(k)}\,\Xi_{+}(w_{1},w_{2})
   + A_{-}^{(k)}\,\Xi_{-}(w_{1},w_{2}),
   \label{eq:Xi_phys_Apm_app}
\end{equation}
we immediately identify
\begin{equation}
  A_{+}^{(k)} = |\braket{k|\chi_{+}}|^{2},
  \qquad
  A_{-}^{(k)} = |\braket{k|\chi_{-}}|^{2},
  \label{eq:Apm_prob_app}
\end{equation}
so that $A_{+}^{(k)}$ and $A_{-}^{(k)}$ are the probabilities to find the
channel $k$ in the geometric eigenmodes $\ket{\chi_{+}}$ and
$\ket{\chi_{-}}$, respectively. As a consequence,
\begin{equation}
  A_{+}^{(k)} + A_{-}^{(k)} = 1.
  \label{eq:Apm_sum_app}
\end{equation}

\subsection{Explicit expression in the Bloch representation}

The coefficients $A_{\pm}^{(k)}$ can be made fully explicit by exploiting the
Bloch representation of SU(2) states. The pure state $\ket{k}$ is described by
a unit Bloch vector $\vec{s}_{k}$, with density matrix
\begin{equation}
  \rho_{k}
  :=
  \ket{k}\!\bra{k}
  = \frac{1}{2}\Big(\mathds{1} + \vec{s}_{k}\!\cdot\!\vec{\sigma}\Big),
  \qquad
  |\vec{s}_{k}| = 1.
  \label{eq:rho_k_app}
\end{equation}
Similarly, the projectors onto the eigenvectors $\ket{\chi_{\pm}}$ of
$\hat{n}\!\cdot\!\vec{\sigma}$ are
\begin{equation}
  P_{\pm}
  := \ket{\chi_{\pm}}\!\bra{\chi_{\pm}}
  = \frac{1}{2}\Big(\mathds{1} \pm \hat{n}\!\cdot\!\vec{\sigma}\Big).
  \label{eq:Ppm_app}
\end{equation}
Using Eq.~\eqref{eq:Apm_prob_app} we can write
\begin{equation}
  A_{\pm}^{(k)}
  = \bra{k}P_{\pm}\ket{k}
  = \mathrm{Tr}\!\big(\rho_{k}\,P_{\pm}\big).
  \label{eq:Apm_trace_app}
\end{equation}
Substituting \eqref{eq:rho_k_app} and \eqref{eq:Ppm_app} and using the Pauli
algebra,
\begin{equation}
  \mathrm{Tr}(\mathds{1})=2,\qquad
  \mathrm{Tr}(\sigma_{i})=0,\qquad
  \sigma_{i}\sigma_{j}=\delta_{ij}\,\mathds{1}
  + i\epsilon_{ijk}\sigma_{k},
  \label{eq:pauli_identities_app}
\end{equation}
one finds
\begin{align}
  A_{\pm}^{(k)}
  &= \mathrm{Tr}\!\left[
      \frac{1}{2}(\mathds{1} + \vec{s}_{k}\!\cdot\!\vec{\sigma})\,
      \frac{1}{2}(\mathds{1} \pm \hat{n}\!\cdot\!\vec{\sigma})
     \right]
  \nonumber\\[2mm]
  &= \frac{1}{4}\,\mathrm{Tr}\!\left[
      \mathds{1}
      \;\pm\;\hat{n}\!\cdot\!\vec{\sigma}
      \;+\;\vec{s}_{k}\!\cdot\!\vec{\sigma}
      \;\pm\;(\vec{s}_{k}\!\cdot\!\vec{\sigma})(\hat{n}\!\cdot\!\vec{\sigma})
     \right]
  \nonumber\\[2mm]
  &= \frac{1}{4}\Big[2 \pm 2\,\vec{s}_{k}\!\cdot\!\hat{n}\Big]
   = \frac{1}{2}\Big(1 \pm \vec{s}_{k}\!\cdot\!\hat{n}\Big).
  \label{eq:Apm_final_app}
\end{align}
Using the definition \eqref{eq:n_hat_app}, this can be written equivalently as
\begin{equation}
  A_{\pm}^{(k)}(w_{1},w_{2})
  = \frac{1}{2}\left[
      1 \pm
      \frac{\vec{s}_{k}\!\cdot\vec{\alpha}(w_{1},w_{2})}
           {|\vec{\alpha}(w_{1},w_{2})|}
    \right].
  \label{eq:Apm_alpha_app}
\end{equation}
If $\gamma_{k}(w_{1},w_{2})$ denotes the angle between the Bloch vector
$\vec{s}_{k}$ and $\vec{\alpha}(w_{1},w_{2})$, i.e.
\begin{equation}
  \cos\gamma_{k}
  =
  \frac{\vec{s}_{k}\!\cdot\vec{\alpha}(w_{1},w_{2})}
       {|\vec{\alpha}(w_{1},w_{2})|},
  \label{eq:gamma_def_app}
\end{equation}
then Eq.~\eqref{eq:Apm_alpha_app} becomes
\begin{equation}
  A_{+}^{(k)} = \frac{1}{2}\big(1 + \cos\gamma_{k}\big)
              = \cos^{2}\!\frac{\gamma_{k}}{2},
  \qquad
  A_{-}^{(k)} = \frac{1}{2}\big(1 - \cos\gamma_{k}\big)
              = \sin^{2}\!\frac{\gamma_{k}}{2},
  \label{eq:Apm_angles_app}
\end{equation}
which makes explicit that $A_{\pm}^{(k)}$ are the probabilities to find the
channel $k$ aligned or antialigned with the geometric direction selected by
the holonomy. In particular, all hadronic channels $k$ share the same
geometric eigenmodes $\Xi_{\pm}(w_{1},w_{2})$ and differ only in their
projection coefficients $A_{\pm}^{(k)}$, thereby inducing nontrivial
correlations among the corresponding physical form factors.

\nocite{*} 
\bibliographystyle{JHEP}
\bibliography{ref.bib}

\end{document}